\documentclass[prd,12pt,nofootinbib,preprint]{revtex4}

\usepackage[T1]{fontenc}
\usepackage{amsmath,amssymb}
\usepackage{epsfig}
\usepackage{url}
\usepackage{subfigure}
\usepackage{graphicx}
\usepackage{grffile}
\usepackage[usenames,dvipsnames]{color}
\usepackage{slashed}
\usepackage{array}
\usepackage{multirow}
\usepackage[colorlinks,citecolor=blue]{hyperref}
\usepackage[utf8x]{inputenc}
\usepackage{color}
\usepackage{cancel}
\usepackage{gensymb}

\def\bar {\overline}

\def\be {\begin{equation}}
\def\ee {\end{equation}}
\def\beq {\begin{equation}}
\def\eeq {\end{equation}}
\def\bea {\begin{eqnarray}}
\def\eea {\end{eqnarray}}

\newcommand{\besub}{\begin{subequations}}
\newcommand{\eesub}{\end{subequations}}

\def\beq{\begin{equation}}
\def\eeq{\end{equation}}
\def\barr{\begin{array}}
\def\earr{\end{array}}

\begin{document}
\title{Resonant leptogenesis in (2,2) inverse see-saw realisation}

\author{Indrani Chakraborty}
\email{indranic@iitk.ac.in}
\affiliation{Department of Physics, Indian Institute of Technology Kanpur, Kanpur, Uttar Pradesh-208016, India} 

\author{Himadri Roy}
\email{himadrir@iitk.ac.in}
\affiliation{Department of Physics, Indian Institute of Technology Kanpur, Kanpur, Uttar Pradesh-208016, India}

\author{Tripurari Srivastava}
\email{tripurarisri022@gmail.com}
\affiliation{Department of Physics and Astrophysics, University of Delhi, Delhi 110007, India}

\begin{abstract} 
In this present work we uphold the standard model (SM) augmented with two right-handed (RH) neutrinos along with two singlet neutral fermions to generate active neutrino masses via (2,2) inverse see-saw mechanism. All entries of the neutrino mass matrix are taken to be complex to make this study a general one. We also investigate if the parameter points compatible with the neutrino oscillation data simultaneously satisfy the experimental bounds coming from the lepton flavour violating (LFV) decays : $\mu \to e \gamma,~ \tau \to e \gamma, ~ \tau \to \mu \gamma$. This study also explores the prospect of producing the baryon asymmetry of the universe through resonant leptogenesis. Here the resonant leptogenesis is induced by the lightest pair of  degenerate mass eigenstates. Upon solving the coupled Boltzmann equations, one can yield a multi-dimensional model parameter space, where the parameter points are compatible with the neutrino oscillation data, constraints coming from the LFV decays and last but not the least, the observed baryon asymmetry of the universe. 
\end{abstract} 
\maketitle

\section{Introduction} 
\label{Intro}
The discovery of a massive neutral scalar Higgs at the Large Hadron Collider~\cite{Aad:2012tfa,Chatrchyan:2012ufa} confirm the mechanism to generate the mass of gauge bosons and fermions, which also leads to the completion of the particle content of the Standard Model (SM). Augmented with several shortcomings, SM is merely termed as an effective theory valid up to a certain scale. One such drawback involves the generation of neutrino mass within SM. In absence of counterpart of left-handed neutrinos in SM, the neutrinos remain massless. Since the observations of neutrino oscillations confirm massive neutrinos, we indeed require a theory beyond SM (BSM) to incorporate them. The simplest and economical remedy for this is to extend the particle content of SM  which can explain the generation of tiny neutrino mass via {\em see-saw mechanism}. There are various types of see-saw mechanisms proposed, depending on the additional BSM particles and their representations, e.g., Type-I~\cite{Minkowski:1977sc,Mohapatra:1979ia,Schechter:1980gr,Schechter:1981cv}, Type-II~\cite{Konetschny:1977bn,Cheng:1980qt,Lazarides:1980nt,Mohapatra:1980yp,Antusch:2007km,Barbieri:1979ag,Magg:1980ut,Felipe:2013kk,Chakraborty:2019uxk,Rodejohann:2004cg,Chen:2010uc,Parida:2020sng}, Type-III~\cite{Foot:1988aq,Albright:2003xb,Suematsu:2019kst,Parida:2016asc,Biswas:2019ygr}. From the oscillation data, these additional particles must be heavy to explain the smallness of active neutrino mass in these simplest see-saw models. Therefore, it becomes difficult to search for these heavy particles in the present experiments. Therefore, low-scale see-saw mechanism can be very interesting to be explored. The inverse see-saw mechanism~\cite{Mohapatra:1986aw,Mohapatra:1986bd,Bernabeu:1987gr,Gavela:2009cd,Parida:2010wq,Garayoa:2006xs,Abada:2014vea,Law:2013gma,Nguyen:2020ehj,Deppisch:2004fa,Arina:2008bb,Dev:2009aw,Malinsky:2009df,Hirsch:2009ra,Blanchet:2010kw,Dias:2012xp,Agashe:2018cuf,Gautam:2020wsd,Zhang:2021olk} is one of the popular ways of generating tiny neutrino mass with TeV scale heavy neutrinos, which can be produced in colliders.  

Besides the aforementioned problem of neutrino mass generation, observed disparity in the baryon and anti-baryon could also not be explained within the framework of the SM. For dynamic generation of baryon asymmetry, the Sakharov conditions needs to be satisfied \cite{Sakharov:1967dj}, which include : (i) baryon number violation, (ii) $C$ or $CP$-violation, (iii) out-of-equilibrium condition. The present value of baryon number asymmetry per unit photon density is~\cite{Aghanim:2018eyx}:

\bea
\eta_B = \frac{n_B - n_{\bar{B}}}{n_\gamma} = (6.12 \pm 0.04) \times 10^{-10} \,.
\label{baryon-limit}
\eea



Complex Yukawa couplings being the source of additional $CP$ violation, {\em lepton asymmetry} can be produced in out-of-equilibrium  decay of heavy neutrinos, which in turn can be translated to the baryon asymmetry via $B+L$ violating but $B-L$ conserving processes due to the SM sphaleron interactions~\cite{Rubakov:1996vz,PhysRevD.30.2212,PhysRevD.28.2019}. Thus one can compensate for the insufficient baryon asymmetry via the aforementioned mechanism called {\em leptogenesis} \cite{Fukugita:1986hr,Covi:1996wh,Roulet:1997xa,Pilaftsis:1997jf,Buchmuller:2005eh,Chun:2007vh,Kitabayashi:2007bs,Davidson:2008bu,Prieto:2009zz,Suematsu:2011va,AristizabalSierra:2011ab,Hambye:2012fh,Kashiwase:2013uy,Borah:2013bza,Hamada:2015xva,Zhao:2020bzx}. 

Among different variants of leptogenesis, the Type-I  thermal leptogenesis (neutrino mass generated by Type-I see-saw mechanism) has been studied extensively in literature \cite{Buchmuller:2002rq,Giudice:2003jh,Buchmuller:2004nz,Chakraborty:2019zas,Rahat:2020mio}. Thermal leptogenesis requires very heavy neutrinos, {\em i.e.} the mass of the lightest heavy neutrino should be  $>10^{9}$ GeV according to {\em Davidson-Ibarra bound} \cite{Davidson:2008bu,Davidson:2002qv}. Therefore the heavy neutrinos are too heavy to be detected in the collider and direct detection experiments. One of the attractive ways of lowering the mass scale of the heavy neutrinos can be achieved by adding extra gauge singlet neutrinos. Thus even with large Yukawa couplings ($\mathcal{O} \sim 1$), tiny (sub-eV scale) active neutrino masses can be achieved in this particular framework, popularly known as {\em inverse see-saw} (ISS) mechanism.  Further, degeneracy in heavy neutrino masses with the mass-splitting being comparable with their decay width, leads to  low scale resonant leptogenesis~\cite{Hambye:2001eu,Hambye_2002,Pilaftsis:2003gt,Hambye:2004jf,Pilaftsis:2003gt,Pilaftsis:2005rv,Cirigliano:2006nu,Xing:2006ms,Branco:2006hz,DeSimone:2007edo,Iso_2011,Iso_2014,Dev_2014,Aoki:2015owa,Dev:2017wwc,Asaka_2019,Brivio_2019,Brdar:2019iem,Mohanty:2019drv,Fong:2021tqj}.  

Based on this idea, we construct a minimal framework to explain the neutrino masses and mixing as well as the observed baryon asymmetry of the universe. For the analysis, we have chosen an economical framework popularly termed as ISS(2,2), {\em i.e.} (2,2) inverse see-saw realisation \cite{Abada:2014vea}. Here the SM is augmented with two generations of singlet right-handed (RH) neutrinos and two generations of singlet neutral fermions. The addition of extra singlet fermions along with RH neutrinos provide a specific structure of $7 \times 7$ neutrino mass matrix which can be written as:

\bea
M_\nu = \begin{pmatrix}
0 & M_D & 0 \\

M_D^T & M_R & M_S \\

0 & M_S^T & \mu 
\end{pmatrix} 
\eea

As will be discussed later, diagonalisation of $M_\nu$ with suitable assumptions leads to TeV scale see-saw which explains the tiny neutrino masses owing to the double suppression by the scale $M_S$ and smallness of $\mu$. Upon diaogonalising $M_\nu$, we can compute the linear combination of three active neutrinos, two RH neutrinos and two singlet fermions to have seven mass eigenstates, three of them being the active ones and the rest of them being heavy neutrinos. Among the heavy modes, there exist two pairs of almost mass degenerate neutrinos. Here the lepton asymmetry is generated from the out-of-equilibrium decay of the lightest mass degenerate pair leading to resonant leptogenesis at TeV scale. Since the heavy neutrinos have additional contribution to lepton flavor violating (LFV) decay $l_i\to l_j^{’} \gamma$, we also investigate the effect of LFV constraints on the model parameter space. A few allied studies could be found in this minimal ISS(2,2) framework itself or in very similar framework constituting a subset of the whole particle spectrum earlier \cite{Blanchet:2010kw,Agashe:2018cuf,Gautam:2020wsd}, where the authors have taken some approximations like purely diagonal  or purely off-diagonal $\mu$ matrix with real entries etc. to ease out the job of diagonalising the neutrino mass matrix analytically. In our study, all the entries of matrix $M_\nu$, {\em i.e.} elements of $M_D, M_S, \mu$ matrix, are considered to be complex to make the analysis more general one. In addition it helps to probe a larger portion of the multi-dimensional model parameter space which remained unexplored with the aforementioned simplifying assumptions. 

The paper is structured as follows. In section [\ref{Model}], we describe the particle content and the interactions. The structure of the neutrino mass matrix along with its relation with neutrino oscillation parameters have been discussed in the same section. Section [\ref{data-fit}], elaborates on the detail of fitting of neutrino oscillation data. In section [\ref{sec:lfv}], the effects of constraints coming from the LFV decays on the parameter space have been discussed. In section [\ref{sec:lepto}], calculation of $CP$ asymmetry along with the solution of coupled Bolzmann equations are presented. In section [\ref{numerical}], we discuss the analysis and results. Finally, we summarise and conclude in section [\ref{conclusion}]. Some of the important formulae have been relegated to Appendix \ref{app:A}, \ref{app:B} and  \ref{app:C}.

\section{Model}
\label{Model}
In this work, we extend the SM minimally by two right-handed neutrinos $N_{R_1}, N_{R_2}$ and two singlet fermions $S_1, S_2$ to generate the neutrino mass and mixings through inverse see-saw mechanism \cite{Mohapatra:1986bd,Mohapatra:1986aw,Wyler:1982dd,Dias:2012xp,Dev:2009aw,Blanchet:2010kw,Ilakovac:1994kj,Deppisch:2004fa,Arina:2008bb,Malinsky:2009df,Hirsch:2009ra,Agashe:2018cuf,Gautam:2020wsd}. $SU(3)_C, SU(2)_L, U(1)_Y$ quantum numbers assigned to the fields can be found in Table \ref{quantum_no}. \footnote{The hyper-charge $Y$ is computed as : $Q = T_3 + \frac{Y}{2}$, where $T_3$ and $Q$ are the weak isospin and electric charge respectively.} In Table \ref{quantum_no}, $\phi$ is the SM Higgs doublet with hyper-charge $Y= + 1$,  $Q_{L_i}, L_{L_i}$ are the left-handed SM quark and lepton doublet respectively. $u_{R_i}, d_{R_i}, \ell_{R_i}$ are right-handed up-type, down-type quark and lepton singlets respectively. 
\begin{table}[htpb!]
\begin{center}
\begin{tabular}{|c|c|c|c|}
\hline
\hspace{5mm} Particles \hspace{5mm} &  \hspace{5mm} $SU(3)_C$ \hspace{5mm} & \hspace{5mm} $SU(2)_L$ \hspace{5mm} &  \hspace{5mm} $U(1)_Y$ ~~\hspace{5mm}\\ \hline \hline
$\phi$ & 1 & 2 & 1 \\ \hline
$Q_{L_i} = \begin{pmatrix}
u_{L_i}\\
d_{L_i}
\end{pmatrix}, ~ i=3$  & 3 & 2 & $\frac{1}{3}$ \\ \hline

$u_{R_i}, ~ i=3$ & 3 & 1 & $\frac{4}{3}$\\
\hline 
$d_{R_i}, ~ i=3$ & 3 & 1 & -$\frac{2}{3}$\\ \hline 
$L_{L_i} = \begin{pmatrix}
\nu_{L_i}\\
\ell_{L_i}
\end{pmatrix}, ~ i=3$ & 1 & 2 & -1\\
\hline 
$\ell_{R_i}, ~ i=3$ & 1 & 1 & -2\\ \hline 
$N_{R_j}, ~ j=2$ & 1 & 1 & 0 \\
\hline 
$S_j, ~ j=2$ & 1 & 1 & 0 \\ \hline
\end{tabular}
\end{center}
\caption{Different quantum number assigned to the particles.}
\label{quantum_no}
\end{table}

The relevant Lagrangian for inverse see-saw mechanism is written as \footnote{$\alpha$ represents flavour of leptons (not to be confused with the Fine structure constant mentioned later).},
\bea
-\mathcal{L}_y = y_{i \alpha} \bar{N}_{R_i} \phi^\dag \ell_{L_\alpha} + \frac{1}{2} M_{R_{ij}} N_{R_i}^T C^{-1} N_{R_j}+ M_{S_{ij}} N_{R_i}^T C^{-1} S_j + \frac{1}{2} \mu_{ij} S_i^T C^{-1} S_j + {\rm h.c.}
\label{lagrangian_flavour}
\eea
Here $y_{i \alpha}$ is the Yukawa coupling matrix with complex entries in our analysis, $C = i \gamma^2 \gamma^0$ is the charge conjugation matrix. Assigning lepton number $L= +1$ to both $N_{R_j}$ and $S_j$s, the Dirac mass term ( first term in Eq.(\ref{lagrangian_flavour})) becomes lepton number conserving. Whereas the Majorana mass terms for $N_{R_j}$ and $S_j$ (second and fourth terms in Eq.(\ref{lagrangian_flavour})) violate the lepton number by two units.

In the basis $(\nu_L^1, \nu_L^2, \nu_L^3, N_{R_1}^c, N_{R_2}^c, S_1, S_2)^T$, the neutrino mass matrix $M_\nu$ can be written as,
\bea
M_\nu =\begin{pmatrix}
0 & 0 & 0 & {M_D}_{1,1} & {M_D}_{1,2} & 0 & 0 \\
0 & 0 & 0 & {M_D}_{2,1} & {M_D}_{2,2} & 0 & 0 \\
0 & 0 & 0 & {M_D}_{3,1} & {M_D}_{3,2} & 0 & 0 \\
{M_D}_{1,1} & {M_D}_{2,1} & {M_D}_{3,1} & {M_R}_{1,1} & {M_R}_{1,2} & {M_S}_{1,1} & {M_S}_{1,2} \\
{M_D}_{1,2} & {M_D}_{2,2} & {M_D}_{3,2} & {M_R}_{1,2} & {M_R}_{2,2} & {M_S}_{2,1} & {M_S}_{2,2} \\
0 & 0 & 0 & {M_S}_{1,1} & {M_S}_{2,1} & \mu_{1,1} & \mu_{1,2} \\
0 &  0 & 0 & {M_S}_{1,2} & {M_S}_{2,2} & \mu_{1,2} & \mu_{2,2}
\end{pmatrix}
\label{Mnu}
\eea
Here each matrix element of $M_\nu$ is taken to be complex to make the analysis the most general one, and thus can be decomposed into real and imaginary parts as :
\bea
&& {M_D}_{i,j} = {M_D}_{i,j}^R + i ~{M_D}_{i,j}^I ~, ~~{M_R}_{l,m} = {M_R}_{l,m}^R + i ~{M_R}_{l,m}^I ~, \nonumber \\
&& {M_S}_{a,b} = {M_S}_{a,b}^R + i ~{M_S}_{a,b}^I ~, ~~ \mu_{p,q } = \mu_{p,q}^R + i ~\mu_{p,q}^I
\eea
In short, the neutrino mass matrix $M_\nu$ becomes :
\bea
M_\nu = \begin{pmatrix}
0 & M_D & 0 \\
M_D^T & M_R & M_S \\
0 & M_S^T & \mu 
\end{pmatrix} 
\eea
with $M_D = y_{i\alpha} \frac{v}{\sqrt{2}}$. With two generations of $N_{R_j}$ and $S_j$s $M_\nu$ is $7\times7$ dimensional. The individual dimensions of $M_D, M_R, M_S$ and $\mu$ are $3 \times 2, ~ 2 \times 2, ~ 2 \times 2, ~ 2 \times 2$ respectively. 

Considering the mass hierarchy $\mu, M_R << M_D << M_S$  and with the see-saw approximation in the inverse see-saw framework, the effective neutrino mass matrix can be written as \cite{CentellesChulia:2020dfh},

\begin{eqnarray}
m_\nu &=& -\begin{pmatrix}
M_D & 0
\end{pmatrix} \begin{pmatrix}
M_R & M_S \\
M_S^T & \mu
\end{pmatrix}^{-1} \begin{pmatrix}
M_D^T \\
0
\end{pmatrix} \,. \nonumber \\
&=& -M_D ~ (M_R - M_S~ \mu^{-1} M_S^T)^{-1} ~ M_D^T \,.
\label{mass-mat}
\end{eqnarray}
Following the aforementioned see-saw hierarchy, we can neglect $M_R$ in Eq.(\ref{mass-mat}) and approximate the active neutrino mass matrix as \cite{Zhou:2012ds}:

\bea
m_\nu &=&  M_D ~ (M_S^T)^{-1} ~\mu ~M_S^{-1} ~M_D^T , \nonumber \\
&& =  M_D ~M_{\rm mid}^{-1}~ M_D^T , ~~ {\rm with} ~~ M_{\rm mid} =  M_S ~\mu^{-1} M_S^T.
\label{mneutrino}
\eea
Elements of $M_{\rm mid}$ are relegated to Appendix \ref{app:A}. With $M_R = 0$, double suppression by the mass scale $M_S$ along with smallness of $\mu$ yield tiny neutrino mass. We shall provide a short discussion on the phenomenological implication of setting $M_R = 0$ in our analysis at the end of Section \ref{numerical}. Upon diagonalising $m_\nu$ one gets the light neutrino masses by the transformation : 
\bea 
U_{\rm PMNS}^T~ m_\nu ~U_{\rm PMNS} = {\rm diag}(m_1,m_2,m_3) = \hat{m_\nu} \,.
\label{mass_diag}
\eea
 where $m_1, m_2, m_3$ are three light active neutrino masses, $U_{\rm PMNS}$ is the Pontecorvo-Maki-Nakagawa-Sakata matrix (PMNS matrix). $U_{\rm PMNS}$ can be written as :
\bea
U_{\rm PMNS} = \begin{pmatrix}
c_{13}c_{12} & c_{13} s_{12} & s_{13}e^{-i \delta_{\rm CP}} \\
-s_{12}c_{23}-c_{12}s_{23}s_{13}e^{i \delta_{CP}} & c_{12} c_{23}-s_{12} s_{23} s_{13} e^{i \delta_{\rm CP}}& s_{23} c_{13} \\
s_{12} s_{23} - c_{12} c_{23} s_{13}e^{i \delta_{\rm CP}} & -c_{12} s_{23} - s_{12} c_{23} s_{13} e^{i \delta_{\rm CP}} & c_{23} c_{13}
\end{pmatrix},
\label{UPMNS}
\eea
Here $c_{ij} \equiv \cos \theta_{ij} , s_{ij} \equiv \sin \theta_{ij}$ and $\delta_{CP}$ is the $CP$-violating phase.

\section{Neutrino data fitting}
\label{data-fit}
Starting from Eq.(\ref{mneutrino}), one can rewrite $\mu$ in terms of $ m_\nu, M_D, M_S$ as :
\bea
\mu = M_S^T~ M_D^{-1}~ m_\nu ~(M_D^T)^{-1}~ M_S
\label{Mmu-fit}
\eea
From Eq.(\ref{mass_diag}), $m_\nu$ can be expressed in terms of $U_{\rm PMNS}, \hat{m_\nu}$. The elements of the $U_{\rm PMNS}$ matrix in Eq.(\ref{UPMNS}) are constrained from neutrino oscillation data. For our analysis, the parameters are fixed at their central values (for normal hierarchy) \cite{Esteban:2020cvm} \footnote{Here we consider normal hierarchy (NH) among the light neutrinos.} :
\bea
&&\sin^2 \theta_{12} = 0.304 , ~ \sin^2 \theta_{23} = 0.573 , ~ \sin^2 \theta_{13} = 0.02219, \nonumber \\
&&\Delta m_{21}^2 = 7.42 \times 10^{-5} {\rm eV}^2 , ~ \Delta m_{31}^2 = 2.517 \times 10^{-3} {\rm eV}^2 , ~ \delta_{\rm CP} = 197^\circ \,.
\label{central-value}
\eea

Having one massless active neutrino is one of the inevitable features of (2,2) inverse see-saw realisation. Therefore, we consider one of the three light active neutrino masses ($m_1$) to be zero, such that $m_1,m_2,m_3$ satisfy : $ (m_1 + m_2 + m_3) \leq 0.12$ eV \cite{Aghanim:2018eyx,Vagnozzi:2017ovm}. With $M_R = 0$, treating the elements of $M_D, M_S$ as input parameters, one can solve for the real and imaginary parts of the elements of $\mu$ at the left hand side of Eq.(\ref{Mmu-fit}), which in turn makes the entries of $\mu$ consistent with the neutrino data mentioned above.
\section{Lepton flavour violation}
\label{sec:lfv}
The $7 \times 7$ unitary matrix $U$, diagonalising $M_\nu$ in Eq.(\ref{Mnu}) can be defined as,
\bea
U^T M_\nu U = {\rm diag}(m_1,m_2,m_3,M_{\tilde{\Psi_1}},M_{\tilde{\Psi_2}},M_{\tilde{\Psi_3}},M_{\tilde{\Psi_4}} )
\eea
Here $m_1,m_2,m_3$ are three light neutrino masses and $M_{\tilde{\Psi_i}}$, with $i=4$, are masses of the heavy neutrinos. \footnote{Since the mass hierarchy condition $\mu << M_D << M_S$ has to hold for neutrino mass generation through inverse see-saw mechanism and the mass splitting between the pair of $M_{\tilde{\Psi_i}}$, {\em i.e.} $|M_{\tilde{\Psi}_{1(3)}}-M_{\tilde{\Psi}_{2(4)}}| \sim \mu$, the respective pairs become mass degenerate. } Thus light neutrino flavor eigenstates can be expressed as linear combinations of seven mass eigenstates $\nu_1^{'},\nu_2^{'},\nu_3^{'},\tilde{\Psi_1},\tilde{\Psi_2},\tilde{\Psi_3},\tilde{\Psi_4}$ :
\bea
\nu_j = \sum_{c=1}^7 (U)_{jc} f_c \,
\eea
where $f_c$ contains all seven mass eigenstates mentioned above.

\begin{figure}[htpb!]
\includegraphics[scale=0.25]{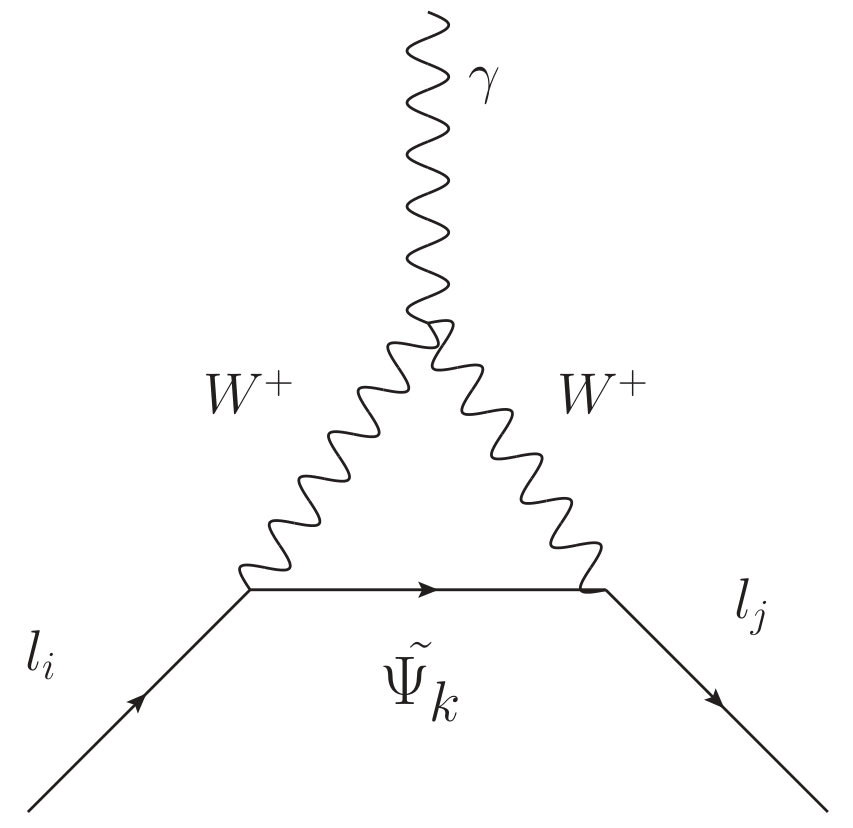}
\caption{Diagram contributing to the LFV decay $l_i \to l_j \gamma$.}
\label{lfv}
\end{figure}

In this scenario, the lepton flavor violating decay $l_i \rightarrow \ l_j \gamma$ obtains additional contributions coming from the heavy neutrinos $\tilde{\Psi_1},\tilde{\Psi_2},\tilde{\Psi_3},\tilde{\Psi_4}$ as shown in Fig.\ref{lfv}.
The branching ratio of the aforementioned LFV process can be computed as \cite{Deppisch:2004fa,Ilakovac:1994kj}:
\bea
{\rm BR} (l_i \rightarrow l_j \gamma) = \frac{\alpha^3 \sin^2 \theta_W}{256 \pi^2}~\left(\frac{M_{l_i}}{M_W}\right)^4~ \frac{M_{l_i}}{\Gamma_{l_i}} ~|G_{ij}|^2
\label{lfv-br}
\eea
with,
\bea
G_{ij} = \sum_k U_{ik}^* ~U_{jk}~ G_\gamma \left(\frac{M_{\tilde{\psi_k}}^2}{M_W^2}\right), ~ G_\gamma(x) = - \frac{2 x^3 + 5 x^2 -x}{4(1-x)^2} - \frac{3 x^3}{2(1-x)^4} ~{\rm ln} x
\eea

Here $\alpha = \frac{e^2}{4 \pi}$ is the fine structure constant, $\theta_W$ is the Weinberg angle, $U_{ij}$ is the $i$-$j$th element of unitary matrix $U$; $M_{l_i}, M_W, M_{\tilde{\Psi_i}}$ are the masses of the decaying lepton, $W$-boson and $\tilde{\Psi_i}$ respectively and $\Gamma_{l_i}$ is the decay width of the decaying lepton. The decay width of $\tau$ is
$\Gamma_\tau = 2.267 \times 10^{-12}$ GeV \cite{10.1093/ptep/ptaa104} and the same for muon is \cite{Ilakovac:1994kj} \footnote{The analytical formula for $\Gamma_\mu$ agrees very well with the experimental data.}:
\bea
\Gamma_\mu = \frac{G_F^2 M_\mu^5}{192 \pi^3}~(1-8 \frac{M_e^2}{M_\mu^2})~\left[1 + \frac{\alpha}{2 \pi} (\frac{25}{4}-\pi^2)\right] \,.
\eea
Here, $G_F, M_\mu, M_e$ are Fermi constant and masses of $\mu$ and $e$ respectively.

The most recent bound on the branching ratios of $l_i \to l_j \gamma$ are listed in Table \ref{lfv:tab}, among which BR$(\mu \to e \gamma)$ is the most constraining \cite{TheMEG:2016wtm}.
  \begin{table}[htpb!]
  \begin{tabular}{|c|c|}
  \hline 
 Branching ratio of LFV process & Experimental upper bound \\ \hline
  \hline 
  BR$(\mu \to e \gamma)$ &  $< 4.2 \times 10^{-13}$ \cite{TheMEG:2016wtm}\\ \hline
  BR$(\tau \to e \gamma)$ &  $< 1.5 \times 10^{-8}$  \cite{Aubert:2009ag}\\ \hline
  BR$(\tau \to \mu \gamma)$ & $< 1.5 \times 10^{-8}$ \cite{Aubert:2009ag} \\ \hline
\hline
 \end{tabular}
	\caption{ Branching ratios of the relevant LFV processes. }
	\label{lfv:tab}
\end{table}	
The effect of the LFV constraints on the parameter space will be discussed in Section \ref{numerical}.
\section{Leptogenesis}
\label{sec:lepto}
\begin{figure}[htpb!]{\centering
\subfigure[]{
\includegraphics[scale=0.15]{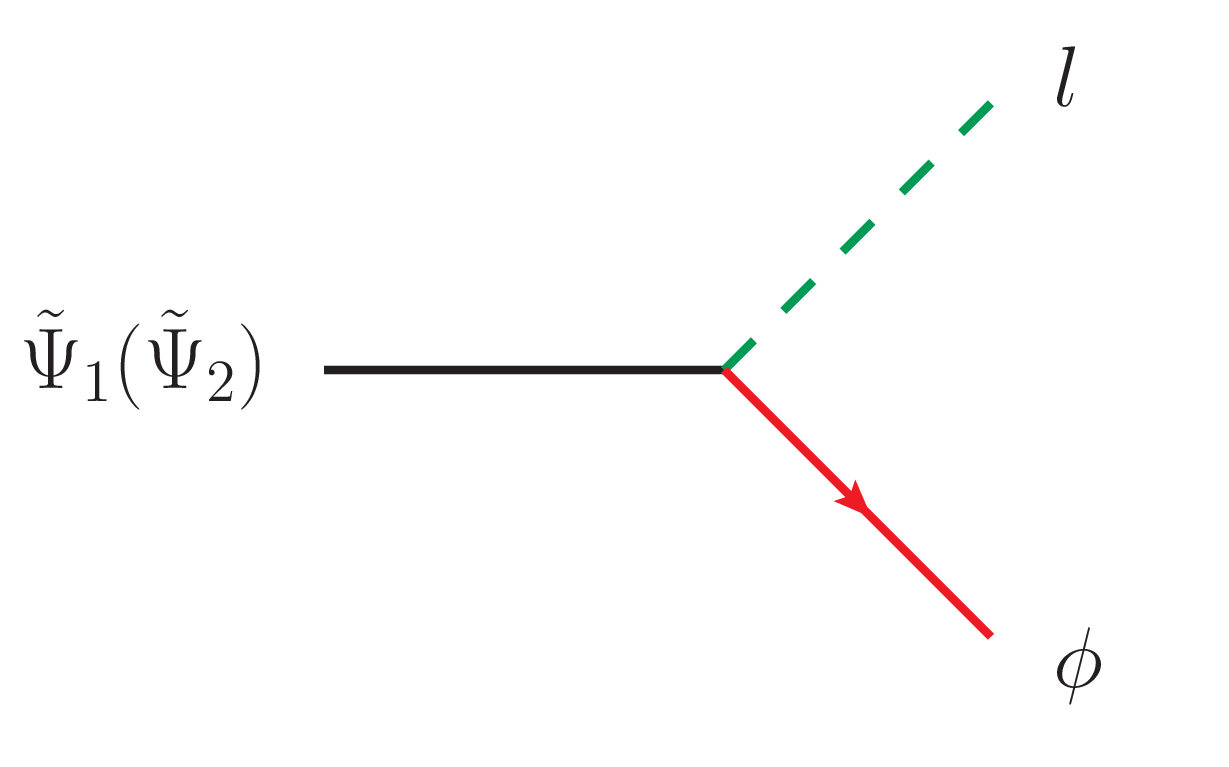}}
\subfigure[]{
\includegraphics[scale=0.15]{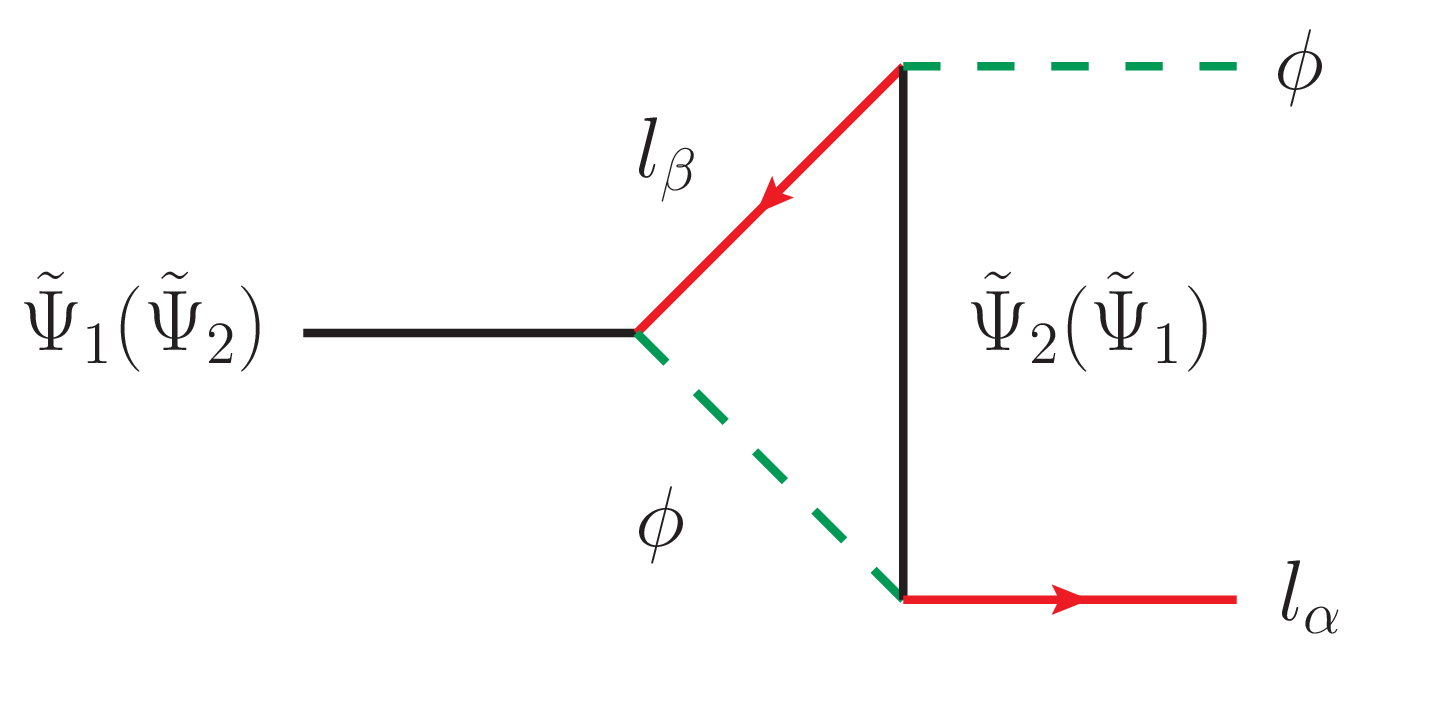}}
\subfigure[]{
\includegraphics[scale=0.15]{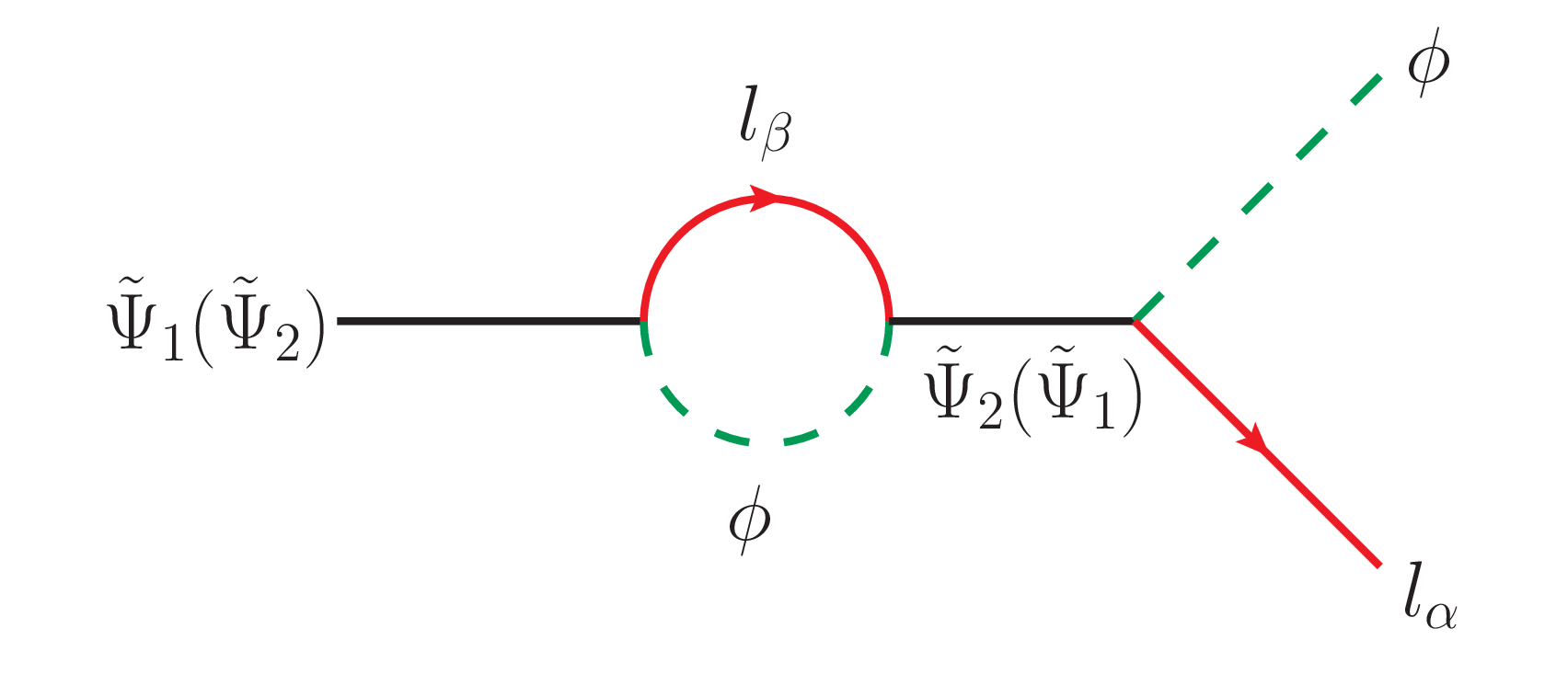}}} 
\caption{Diagrams contributing to the $CP$-asymmetry $\epsilon_1$ and $\epsilon_2$: (a) tree-level decay of $\tilde{\Psi_{1}} (\tilde{\Psi_{2}})$, (b) vertex correction, (c) self-energy diagram.}
\label{cp-asymmetry}
\end{figure}

In this work, we aim to generate the observed baryon asymmetry through leptogenesis. We are interested to analyse the parameter space consistent with the neutrino data and present baryon asymmetry. The $CP$-asymmetry will be calculated from the out of equilibrium decay of the lightest mass degenerate pair ($\tilde{\Psi_{1}}, \tilde{\Psi_{2}}$) among four aforementioned mass eigenstates $\tilde{\Psi_{1}}, \tilde{\Psi_{2}}, \tilde{\Psi_{3}}, \tilde{\Psi_{4}}$, as shown in Fig.\ref{cp-asymmetry}. With this mass hierarchy among $\tilde{\Psi_{i}}$s, we will be operating within the framework of resonant leptogenesis \cite{Hambye:2001eu,Hambye_2002,Pilaftsis:2003gt,Hambye:2004jf,Pilaftsis:2003gt,Pilaftsis:2005rv,Cirigliano:2006nu,Xing:2006ms,Branco:2006hz,DeSimone:2007edo,Iso_2011,Iso_2014,Dev_2014,Aoki:2015owa,Dev:2017wwc,Asaka_2019,Brivio_2019,Brdar:2019iem,Mohanty:2019drv,Fong:2021tqj} , where the $CP$ -asymmetry is enhanced by considering the mass-splitting
between any two of the heavy neutrinos to be comparable with their decay width.
Thus here we have to apply two flavour approximation for the decay of the lightest pair of degenerate neutrinos into another and solve three coupled Boltzmann equations as will be discussed later. After solving three Boltzmann equations simultaneously, one ends up with three solutions, {\em i.e.} comoving densities 
$Y_{\tilde{\Psi_{1}}}, Y_{\tilde{\Psi_{2}}}$ and $Y_{B-L}$ of $\tilde{\Psi_{1}}, \tilde{\Psi_{2}}$ (assuming $\tilde{\Psi_{1}}, \tilde{\Psi_{2}}$ are almost mass degenerate and lighter than other two states $\tilde{\Psi_{3}}, \tilde{\Psi_{4}}$) and $B-L$ asymmetry respectively. Here the comoving density is defined as the ratio of actual number density \footnote{The number densities of particles with mass M and temperature T can be written as :
\bea
N_{eq} = \frac{g M^2 T}{2 \pi^2}K_2(\frac{M}{T})
\eea
$g$ being the number of degrees of freedom of corresponding particles, $K_2$ being second modified Bessel function of second kind.} and the entropy density $\bar{s}$ of the universe \footnote{Entropy density is calculated as : $\bar{s} = \frac{2 \pi^2}{45}~ g_{eff} T^3$. Here $T$ is the temperature and $g_{eff}$ is the number of degrees of freedom (D.O.F), which is calculated in Appendix \ref{app:C}.}.

Next let us proceed to calculate the $CP$-asymmetry produced from the decay of $\tilde{\Psi_{1}} (\tilde{\Psi_{2}})$ into $\tilde{\Psi_{2}} (\tilde{\Psi_{1}})$ and then solve the coupled Boltzmann equations. 
\subsection{Computation of $CP$-asymmetry}

With $M_R = 0$, $M_\nu$ in Eq.(\ref{Mnu}) is block diagonalisable. Upon diagonalising lower block of $M_\nu$ ($4\times 4$ complex symmetric sub-matrix), 

\bea
\mathcal{M} = \begin{pmatrix}
0 & M_S \\
M_S^T & \mu
\end{pmatrix}
\eea

with $4\times 4$ unitary matrix $V$, one ends up with two mass-degenerate pairs ($\tilde{\Psi_1},\tilde{\Psi_2}$) and ($\tilde{\Psi_3},\tilde{\Psi_4}$) as :

\bea
M_{\rm diag} = V^T \mathcal{M} V = {\rm diag} ~(M_{\tilde{\Psi_1}},M_{\tilde{\Psi_2}}, M_{\tilde{\Psi_3}}, M_{\tilde{\Psi_4}})
\eea

As mentioned earlier, the mass splittings between the mass-degenerate pairs are proportional to $\mu$.  Diagonalising the $4 \times 4$ sub-matrix $\mathcal{M}$ with all complex entries in $M_S$ and $\mu$ analytically is a challenging task. Instead we shall diagonalise it numerically in our analysis.

To calculate the $CP$-asymmetry, it is preferable to compute it using that particular basis, where the $\mathcal{M}$ is diagonal. In this basis, the Lagrangian in Eq.(\ref{lagrangian_flavour}) can be restructured as, 
\bea
-\mathcal{L}_h = h_{i \alpha} \bar{\tilde{\Psi_i}} \phi^\dag \ell_{L_\alpha} + \frac{1}{2} M_{\rm diag} \tilde{\Psi_i}^T C^{-1} \tilde{\Psi_i} + {\rm h.c.}
\label{lagrangian_mass}
\eea
Yukawa couplings in the diagonal mass basis ($h_{i \alpha}$) are connected to the Yukawa couplings in the flavour basis ($y_{i \alpha}$) through following relations :
\bea
h_{1 \alpha} &=&  V_{11}^*~ y_{1 \alpha} + V_{12}^*~ y_{2 \alpha} \nonumber \\
h_{2 \alpha} &=&  V_{21}^*~ y_{1 \alpha} + V_{22}^*~ y_{2 \alpha}
\nonumber \\
h_{3 \alpha} &=&  V_{13}^*~ y_{1 \alpha} + V_{23}^*~ y_{2 \alpha}
\nonumber \\
h_{4 \alpha} &=&  V_{14}^*~ y_{1 \alpha} + V_{24}^*~ y_{2 \alpha}
\eea
For the decay of $\tilde{\Psi_j}$ into $\ell_\alpha \phi~ (\bar{\ell_\alpha} \phi^\dag)$, one can compute the total $CP$-asymmetry $\epsilon_j$ by summing over the SM flavour $\alpha$,
\bea
\epsilon_j = \frac{\sum_\alpha \left[ \Gamma(\tilde{\Psi_i} \rightarrow \ell_\alpha \phi) - \Gamma(\tilde{\Psi_i} \rightarrow \bar{\ell}_\alpha \phi^\dag) \right]}{\sum_\alpha \left[ \Gamma(\tilde{\Psi_i} \rightarrow \ell_\alpha \phi) + \Gamma(\tilde{\Psi_i} \rightarrow \bar{\ell}_\alpha \phi^\dag) \right]} = \frac{1}{8 \pi} \sum_{j\neq i} \frac{{\rm Im}[(h h^\dag)_{ij}^2]}{(h h ^\dag)_{ii}} f_{ij}
\eea
Here $f_{ij}$ receives contributions both from the vertex correction and self energy correction. As we are considering the decays of pseudo mass degenerate states, the leptogenesis will be dominantly of {\em resonant} type.  For resonant leptogenesis $f_{ij} \sim f_{ij}^{\rm self}$, with $f_{ij}^{\rm self} = \frac{(M_i^2 - M_j^2) M_i M_j}{(M_i^2 - M_j^2)^2 + R_{ij}^2}$. Following \cite{Garny:2011hg,Iso:2014afa,Iso:2013lba}, here we shall consider $R_{ij} = |M_i \Gamma_i + M_j \Gamma_j|$  {\footnote{In other references, different forms of the regulator such as $R_{ij} = M_i \Gamma_j$ \cite{Pilaftsis:2003gt,Pilaftsis:1998pd,Pilaftsis:1997dr}, $|M_i \Gamma_i - M_j \Gamma_j|$ \cite{Buchmuller:1997yu} can be found. Following the Kadanoff-Baym approach to the resonant leptogenesis and considering the "off-shell contributions", one can extract the CP-asymmetry parameter $\epsilon$ and the correct from of regulator comes out to be $|M_i \Gamma_i + M_j \Gamma_j|$ \cite{Garny:2011hg,Iso:2013lba}. Thus we have used this correct form of regulator to evaluate the accurate lepton asymmetry. Following this treatment, we obtained $\epsilon_1, \epsilon_2 \sim 10^{-7}-10^{-8}$.}}, $\Gamma_j = \frac{(h h^\dag)_{ii} M_j}{8 \pi}$ being the total decay width of $\tilde{\Psi_j}$. 

Thus from the decays of two pseudo degenerate pairs $\tilde{\Psi_1},~\tilde{\Psi_2}$ and $\tilde{\Psi_3},~\tilde{\Psi_4}$, one can compute the $CP$-asymmetry as,
\bea
\epsilon_1 &=& \frac{1}{8\pi (h h^\dag)_{11}} {\rm Im} [(h h^\dag)^2_{12} f_{12} + (h h^\dag)^2_{13} f_{13} + (h h^\dag)^2_{14} f_{14}] \, \nonumber \\
\epsilon_2 &=& \frac{1}{8\pi (h h^\dag)_{22}} {\rm Im} [(h h^\dag)^2_{21} f_{21} + (h h^\dag)^2_{23} f_{23} + (h h^\dag)^2_{24} f_{24}] \, \nonumber \\
\epsilon_3 &=& \frac{1}{8\pi (h h^\dag)_{33}} {\rm Im} [(h h^\dag)^2_{31} f_{31} + (h h^\dag)^2_{32} f_{32} + (h h^\dag)^2_{34} f_{34}] \, \nonumber \\
\epsilon_4 &=& \frac{1}{8\pi (h h^\dag)_{44}} {\rm Im} [(h h^\dag)^2_{41} f_{41} + (h h^\dag)^2_{42} f_{42} + (h h^\dag)^2_{43} f_{43}] \,
\eea 
Since the lepton asymmetry produced in the decay of the heavier pair will be abolished by the lepton number violating scattering of the lighter pair as shown in Fig.\ref{diagram}, we shall consider the decay of the lightest pair among the two for computing the $CP$-asymmetry. For example, if $\tilde{\Psi_1},~\tilde{\Psi_2}$ are lighter than $\tilde{\Psi_3},~\tilde{\Psi_4}$ with $M_{\tilde{\Psi_1}} \sim M_{\tilde{\Psi_2}}$, one can consider the decay of $\tilde{\Psi_1} (\tilde{\Psi_2})$ into $\tilde{\Psi_2} (\tilde{\Psi_1})$ (Fig.\ref{cp-asymmetry}) and calculate the $CP$ asymmetry $\epsilon_1$ and $\epsilon_2$, which will enter in to the Boltzmann equations.
\subsection{Solving Boltzmann equations}
In general, the Boltzmann equations for $\tilde{\Psi_{1}}, \tilde{\Psi_{2}}$ and  the $(B-L)$ asymmetry can be written as \cite{Plumacher:1996kc},
 \bea
 \frac{\text{d} Y_{\tilde{\Psi_{1}}}}{\text{d} z} &=& - \frac{z}{s \hspace{1mm} H(M_{\tilde{\Psi_1}})} \Big[ \Big(\frac{Y_{\tilde{\Psi_{1}}}}{Y^{eq}_{\tilde{\Psi_{1}}}} - 1\Big)(\gamma_{D}^{(1)} + 2 \gamma^{(1)}_{\phi,s} + 4 \gamma^{(1)}_{\phi,t})\Big] \,,
  \label{boltzmann-eq1}
 \eea
  \bea
 \frac{\text{d} Y_{\tilde{\Psi_{2}}}}{\text{d} z} &=& - \frac{z}{s \hspace{1mm} H(M_{\tilde{\Psi_1}})} \Big[ \Big(\frac{Y_{\tilde{\Psi_{2}}}}{Y^{eq}_{\tilde{\Psi_{2}}}} - 1\Big)(\gamma_{D}^{(2)} + 2 \gamma^{(2)}_{\phi,s} + 4 \gamma^{(2)}_{\phi,t})\Big] \,,
  \label{boltzmann-eq2}
 \eea
  \bea
 \frac{\text{d} Y_{B-L}}{\text{d} z}  &=& - \frac{z}{s \hspace{1mm} H(M_{\tilde{\Psi_1}})} \Big[ \sum_{j=1}^2 \left\{\frac{1}{2} \frac{Y_{B-L}}{Y^{eq}_{l}} + \epsilon_j ~\Big(\frac{Y_{\tilde{\Psi_{j}}}}{Y^{eq}_{\tilde{\Psi_{j}}}} - 1\Big) \right\} \gamma_{D}^{(j)}  \nonumber \\
 && +\frac{Y_{B-L}}{Y^{eq}_{l}}\left\{2 \gamma_{\tilde{\Psi},s} + 2 \gamma_{\tilde{\Psi},t}\right\}+ \frac{Y_{B-L}}{Y^{eq}_{l}} \sum_{j=1}^2 \left\{ 2 \gamma^{(j)}_{\phi,t} + \frac{Y_{\tilde{\Psi_{j}}}}{Y^{eq}_{\tilde{\Psi_{j}}}}  \gamma^{(j)}_{\phi,s} \right\} \Big] \,,
 \label{boltzmann-eq3}
 \eea
where $z = \frac{M_{\tilde{\Psi_1}}}{T}$ and $H(M_{\tilde{\Psi_1}})$ is the Hubble parameter at $T = M_{\tilde{\Psi_1}}$ and $H(T=M_{\tilde{\Psi_1}}) = 1.66 ~g_{eff}^{1/2} \frac{T^{2}}{M_{\rm{Pl}}}|_{T=M_{\tilde{\Psi_1}}} $, $M_{\rm{Pl}} = 10^{19}$ GeV being Planck scale. $Y_{\tilde{\Psi_{j}}}^{eq}, Y_l^{eq}$ are the comoving densities at equilibrium. We solve these three equations with initial conditions :
\bea
Y_{\tilde{\Psi_i}}(0) = Y_{\tilde{\Psi_i}}^{eq} , ~{\rm and}~ Y_{B-L}(0) = 0 \,.
\eea
at $T >> M_{\tilde{\Psi_1}}$.

\begin{figure}[htpb!]{\centering
\subfigure[]{
\includegraphics[scale=0.1]{pshi1_l_phi.png}}
\subfigure[]{
\includegraphics[scale=0.1]{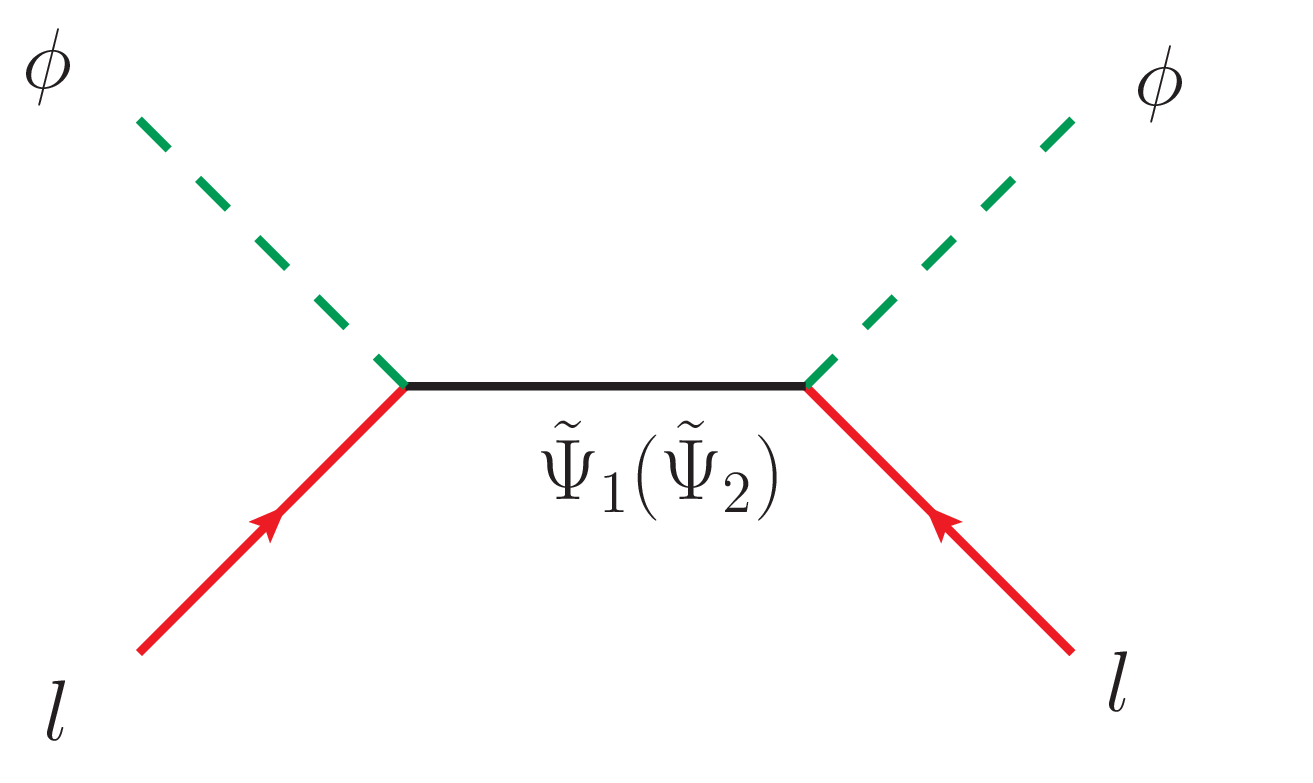}}
\subfigure[]{
\includegraphics[scale=0.1]{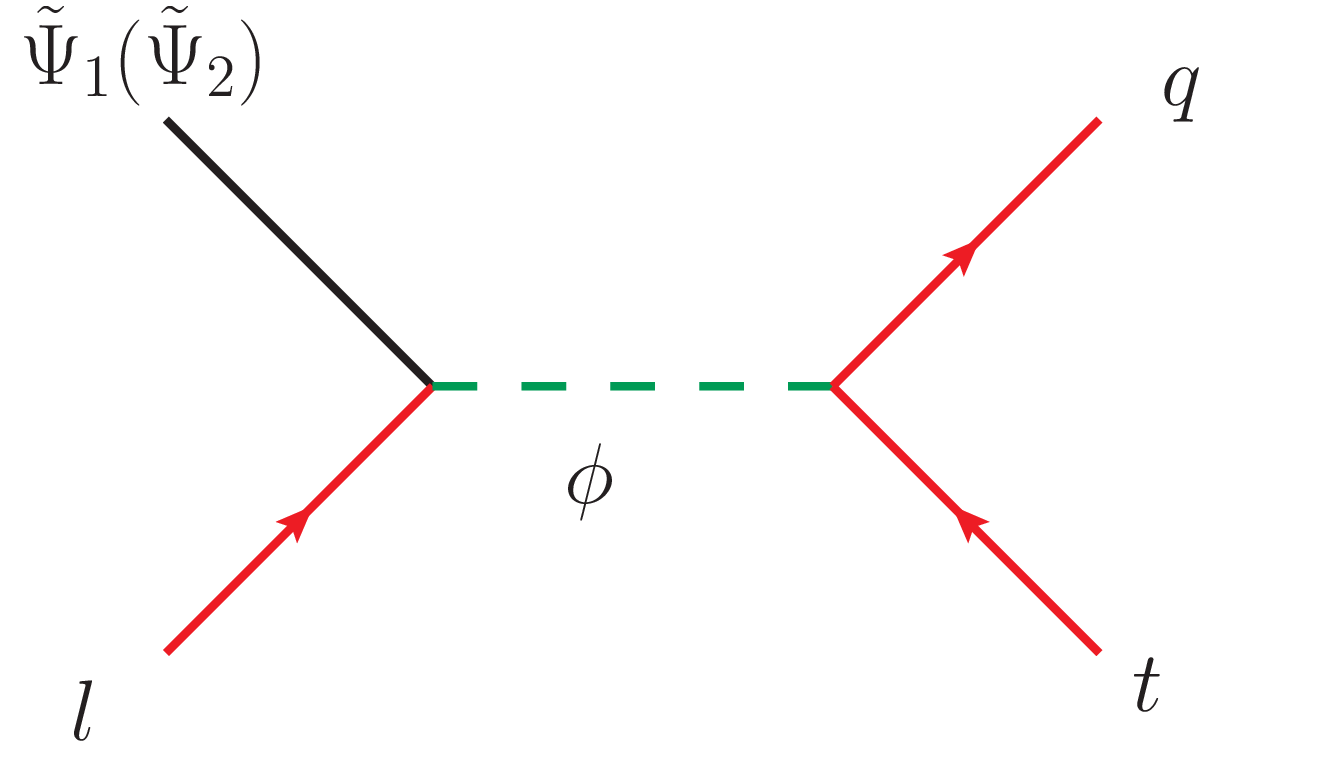}} \\
\subfigure[]{
\includegraphics[scale=0.1]{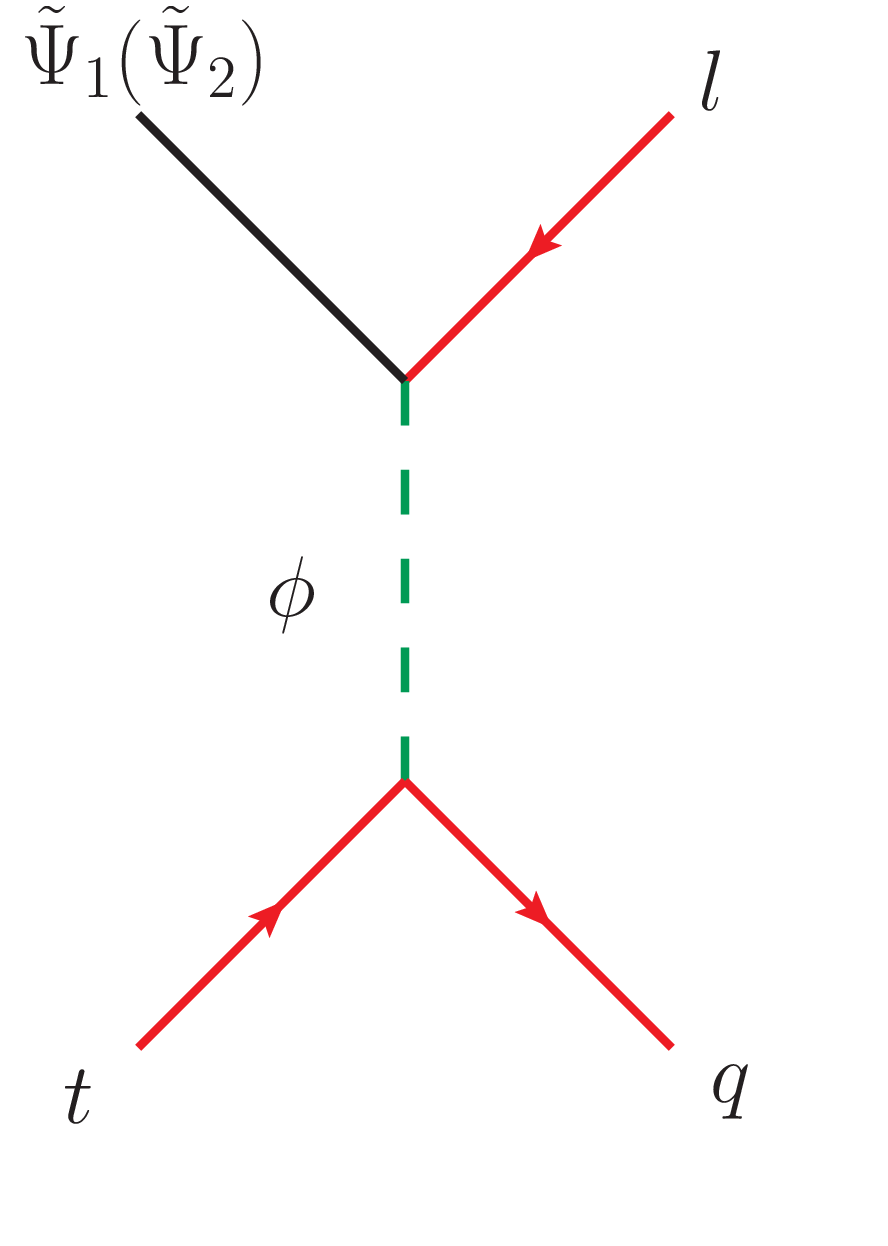}}
\subfigure[]{
\includegraphics[scale=0.1]{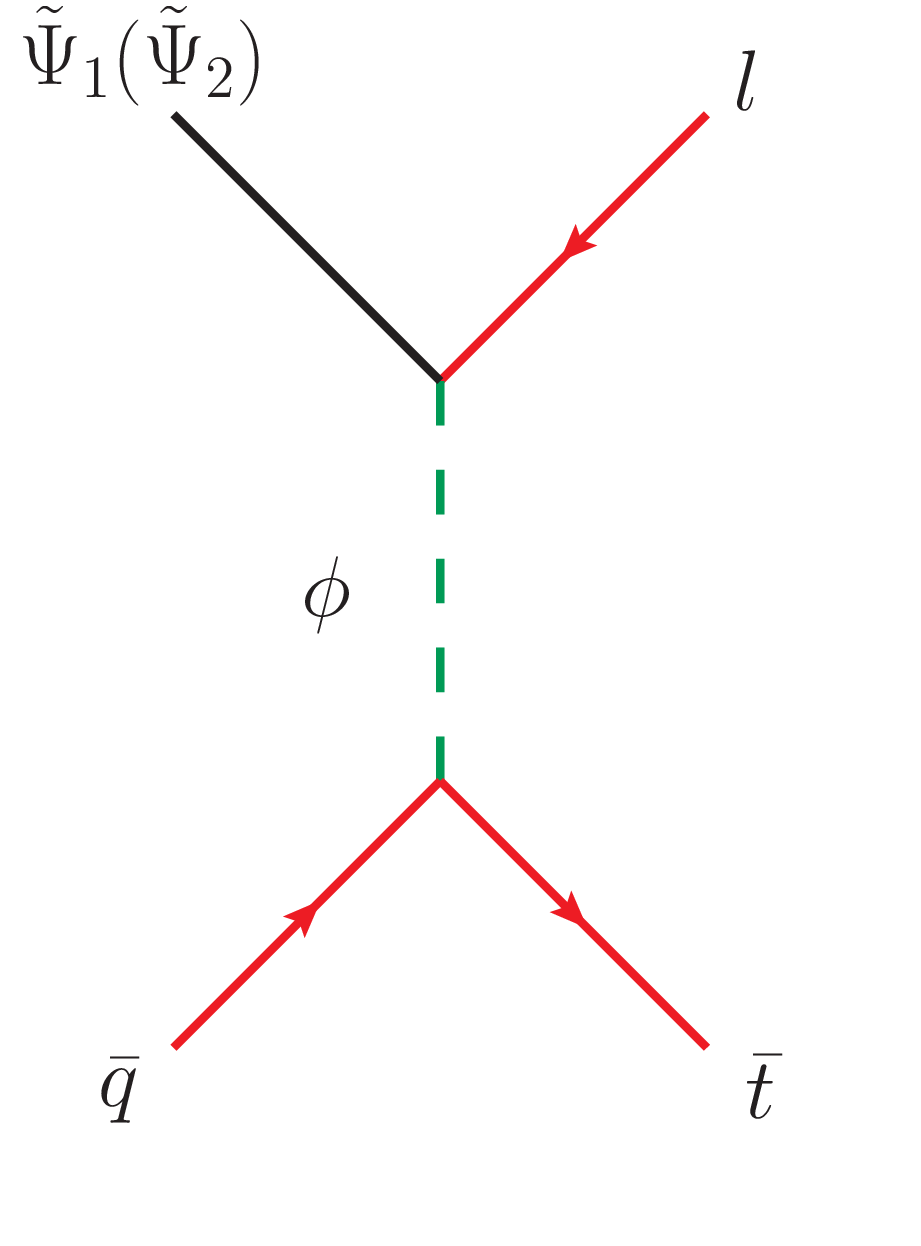}}
\subfigure[]{
\includegraphics[scale=0.1]{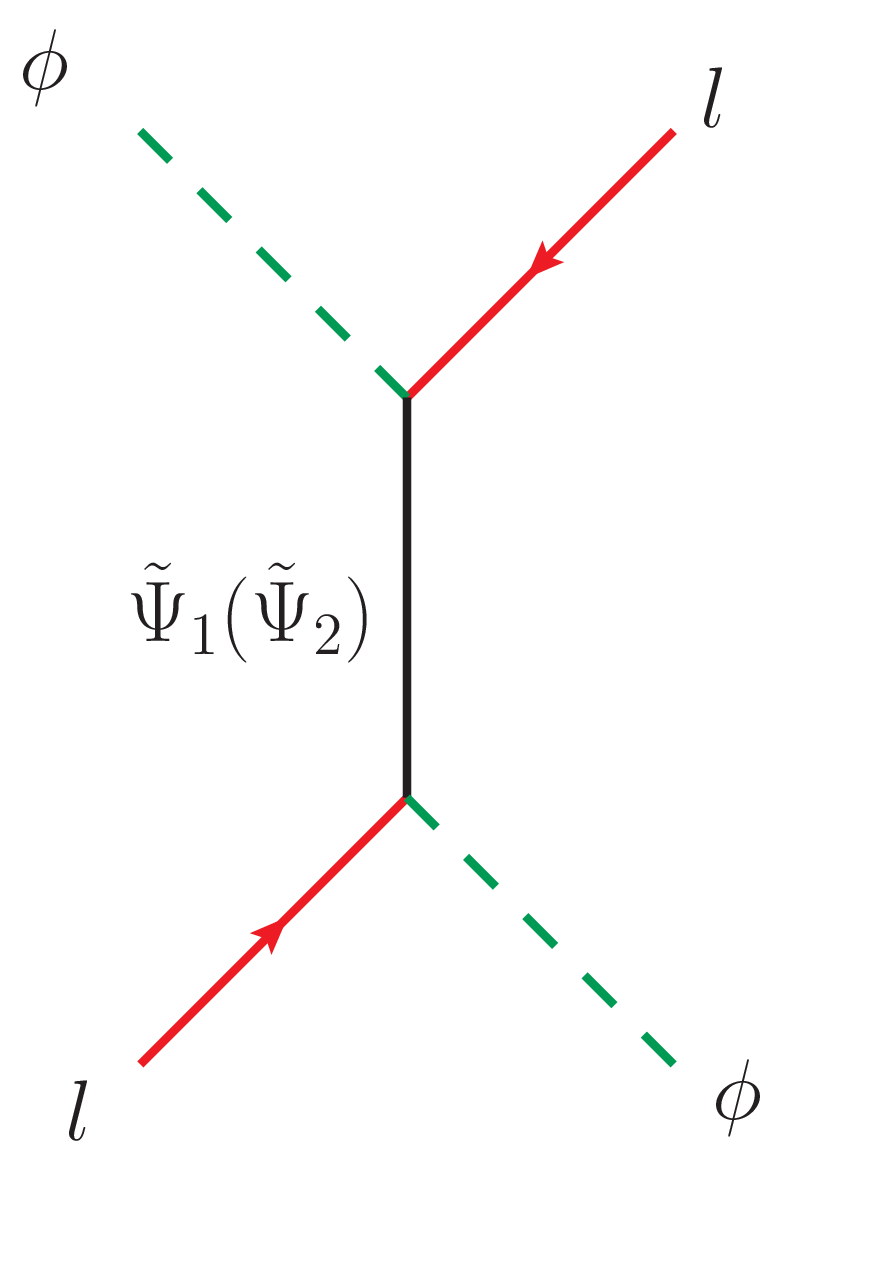}}} 
\caption{(a) Decay of lightest heavy RH-neutrino $\tilde{\Psi_{1}}~(\tilde{\Psi_{2}})$ (contributes to $\gamma_{D}^{(1)}, \gamma_{D}^{(2)}$), (b) $\Delta L = 2$, $s$-channel scattering via $\tilde{\Psi_{1}}~(\tilde{\Psi_{2}})$ (contributes to $\gamma_{\tilde{\Psi},s}$), (c) $\Delta L = 1$, $s$-channel scattering via Higgs (contributes to $\gamma^{1}_{\phi,s}, \gamma^{2}_{\phi,s}$), (d) and (e) $\Delta L = 1$, $t$-channel scattering via Higgs (contributes to $\gamma^{1}_{\phi,t}, \gamma^{2}_{\phi,t}$), (f) $\Delta L = 2$, $t$-channel scattering via $\tilde{\Psi_{1}}~(\tilde{\Psi_{2}})$ (contributes to $\gamma_{\tilde{\Psi},t}$). }
\label{diagram}
\end{figure}
In the first and second Boltzmann equations (Eq.(\ref{boltzmann-eq1}) and Eq.(\ref{boltzmann-eq2})), $\gamma_{D}^{(1)}, \gamma_{D}^{(2)}$ symbolise the contributions from the tree-level decay of $\tilde{\Psi_{1}}, \tilde{\Psi_{2}}$ respectively depicted in Fig.\ref{diagram}(a). All other $\gamma$s in three coupled Boltzmann equations signify space-time densities of different scattering processes mentioned in Fig.\ref{diagram}. Fig.\ref{diagram}(c),(d) and (e) describe $\Delta L=1$ lepton number violating $s$-channel and $t$-channel scalar mediated processes. The contribution from the former processes appears in first two Boltzmann equations as $ \gamma^{j}_{\phi,s}$ and  $\gamma^{j}_{\phi,t}$ with $\tilde{\Psi_{j}}$ in the initial state.
$\Delta L = 2$ lepton number violating $s$-channel and $t$-channel processes mediated by both $\tilde{\Psi_{1}}, \tilde{\Psi_{2}}$ in Fig.\ref{diagram}(b) and Fig.\ref{diagram}(f) yield $\gamma_{\tilde{\Psi},s}$ and $\gamma_{\tilde{\Psi},t}$ respectively in Eq.(\ref{boltzmann-eq3}). Formulae for all the $\gamma$s mentioned in Eq.(\ref{boltzmann-eq1}), (\ref{boltzmann-eq2}) and (\ref{boltzmann-eq3}) are relegated to Appendix \ref{app:B}.

The lepton asymmetry is converted into baryon asymmetry, resulting in the final baryon number at $T_{\rm sph} \sim 150$ GeV (the freeze-out temperature of the sphelaron process ) as \cite{Burnier:2005hp}:
\bea
Y_{B} = \bigg( \frac{8 N_{f} + 4 N_{H}}{22 N_{f} + 13 N_{H} } \bigg) Y_{B-L} (z_{\rm sph}).
\label{baryon-asym-con}
\eea
With $Y_{B-L} (z_{\rm sph})$ as the solution of Boltzmann equations at $z = z_{\rm sph} = \frac{M_{\tilde{\Psi_1}}}{T_{\rm sph}}$.

Here $N_f$ and $N_H$ are the number of generations of fermion families and number of Higgs doublets respectively. In our scenario, $N_f = 3$ and $N_H =1$.
\section{Numerical results}
\label{numerical}
Let us begin this section with the details of neutrino data fitting. Here we start with the neutrino mass matrix $M_\nu$ in Eq.(\ref{Mnu}), whose elements are taken to be complex to make the analysis a general one.  As mentioned earlier, throughout the analysis we set $M_R = 0$ to bring in a resemblance with the original {\em inverse see-saw} model. Staring from Eq.(\ref{mneutrino}), one can solve for $\mu$ following Eq.(\ref{Mmu-fit}), satisfying the neutrino oscillation data. Expressing $m_\nu$ at the right hand side of Eq.(\ref{Mmu-fit}) in terms of $U_{\rm PMNS}$ and $\hat{m_\nu}$ and considering the central values of the oscillation parameters as in Eq.(\ref{central-value}), one can compute $\mu$ in terms of $M_D, M_S$ and neutrino oscillation parameters with $M_R = 0$. Thus the real and imaginary parts of all the entries of $\mu$ are compatible with the neutrino oscillation data. Since one of the unavoidable features of the (2,2) inverse see-saw realisation is a zero active neutrino mass eigenstate, we consider one of the active neutrino masses to be zero while fitting the neutrino data. With the aforementioned assumption, here we only consider the normal mass hierarchy between the active neutrinos. For a systematic study, let us  divide $M_{\tilde{\Psi_1}}$ in the following ranges : (a) $M_{\tilde{\Psi_1}} < 10$ TeV, (b) 10 TeV $< M_{\tilde{\Psi_1}} <$ 100 TeV, (c) 100 TeV $< M_{\tilde{\Psi_1}} <$ 1000 TeV, (d) $M_{\tilde{\Psi_1}} >$ 1000 TeV. For the aforementioned four mass ranges, let us quote the ranges of the input parameters in Table \ref{input-param}.
  \begin{table}[htpb!]
  \resizebox{17cm}{!}{
  \begin{tabular}{|c|c|c|c|}
  \hline 
  $M_{\tilde{\Psi_1}} < 10$ TeV & 10 TeV $< M_{\tilde{\Psi_1}} <$ 100 TeV & 100 TeV $< M_{\tilde{\Psi_1}} <$ 1000 TeV & $M_{\tilde{\Psi_1}} >$ 1000 TeV \\ \hline 
  \hline 
$ M_{D_{i,j}}^{R,I} \in [10^7~{\rm eV} : 10^8~{\rm eV}]$,  & $ M_{D_{i,j}}^{R,I} \in [10^7~{\rm eV} : 10^8~{\rm eV}]$ & $ M_{D_{i,j}}^{R,I} \in [10^7~{\rm eV} : 10^8~{\rm eV}]$ & $ M_{D_{i,j}}^{R,I} \in [10^7~{\rm eV} : 10^8~{\rm eV}]$ \\ 
 $M_{S_{i,j}}^{R,I} \in [10^{10}~{\rm eV} : 10^{13}~{\rm eV}]$  & $M_{S_{i,j}}^{R,I} \in [10^{12}~{\rm eV} : 10^{14}~{\rm eV}]$ & $M_{S_{i,j}}^{R,I} \in [10^{13}~{\rm eV} : 10^{15}~{\rm eV}]$ & $M_{S_{i,j}}^{R,I} \in [10^{14}~{\rm eV} : 10^{16}~{\rm eV}]$\\ \hline
\hline
 \end{tabular}}
	\caption{ Values of the input parameters for four mass regimes. }
	\label{input-param}
\end{table}	

Having obtained the parameter space compatible with the neutrino oscillation data, next we compute the branching ratios of the lepton flavour violating (LFV) processes like $l_i \to l_j \gamma$ following Eq.(\ref{lfv-br}) for each and every point in the parameter space and check if the model parameters altogether tune themselves to yield correct BR($l_i \to l_j \gamma$), compatible with the experimental data tabulated in Table \ref{lfv:tab}. Among the three processes tabulated in Table \ref{lfv:tab}, the strongest bound originates from $\mu \to e \gamma$ coming from the MEG experiment \cite{TheMEG:2016wtm}. In addition, we have also considered the constraint coming from the observed baryon asymmetry. The total parameter space (compatible with the neutrino oscillation data and experimental LFV branching ratios) has been divided into three parts : (i) the points lying within the correct baryon asymmetry band mentioned in Eq.(\ref{baryon-limit}), (ii) the points which cannot generate adequate baryon asymmetry and thus lie below the lower edge of the aforementioned band (under abundant), (iii) the points corresponding to higher values of baryon asymmetry, which lie above the upper edge of the band (over abundant). Among these three parts, the third part of the parameter space is said to be ruled out by the model parameters, while the first two are allowed. Let us now describe the features of the parameter spaces for four mass regimes under all the aforementioned constraints.

\begin{figure}[htpb!]{\centering
\subfigure[]{
\includegraphics[width=7.5cm,height=5cm, angle=0]{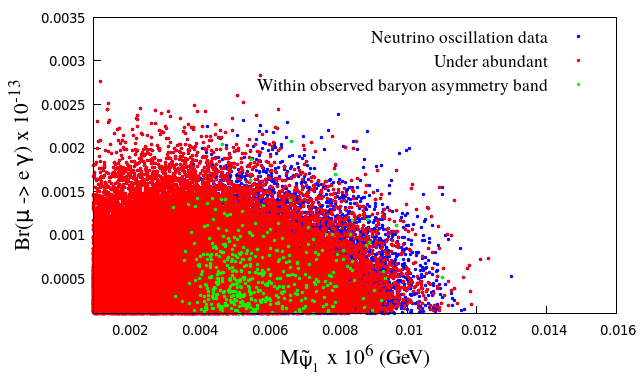}} 
\subfigure[]{
\includegraphics[width=7.5cm,height=5cm, angle=0]{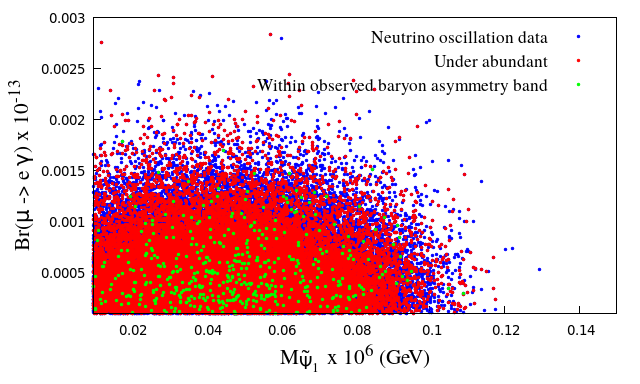}} \\
\subfigure[]{
\includegraphics[width=7.5cm,height=5cm, angle=0]{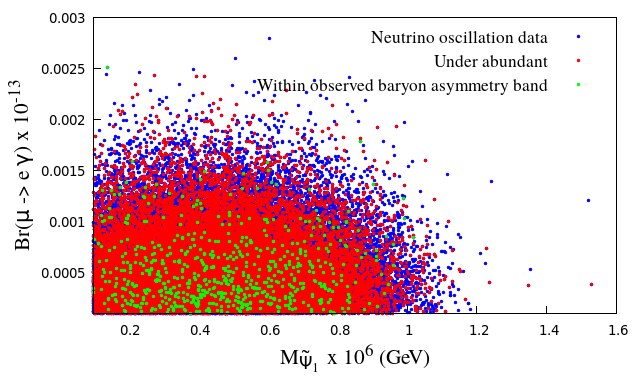}}  
\subfigure[]{
\includegraphics[width=7.5cm,height=5cm, angle=0]{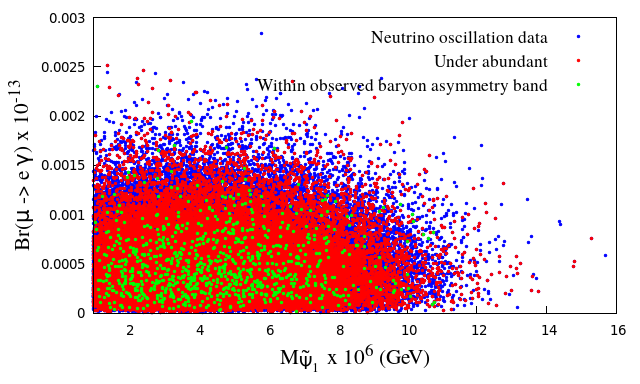}}}
\caption{Variation of BR($\mu \to e \gamma$) with $M_{\tilde{\Psi_1}}$ for (a) $M_{\tilde{\Psi_1}} < 10$ TeV, (upper left) (b) 10 TeV $< M_{\tilde{\Psi_1}} <$ 100 TeV, (upper right) (c) 100 TeV $< M_{\tilde{\Psi_1}} <$ 1000 TeV, (lower left) (d) $M_{\tilde{\Psi_1}} >$ 1000 TeV (lower right). Color coding is expressed in legends.}
\label{Br:m1}
\end{figure}

\begin{figure}[htpb!]{\centering
\subfigure[]{
\includegraphics[width=7.5cm,height=5cm, angle=0]{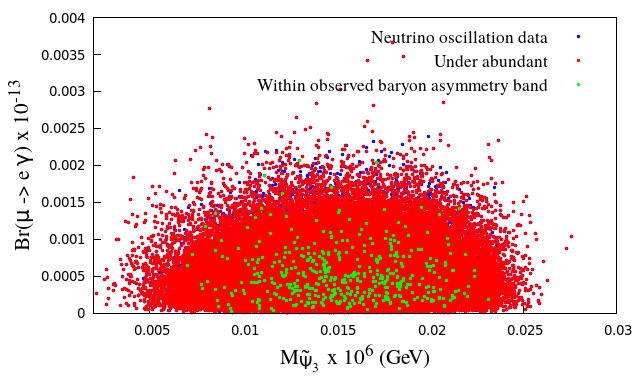}} 
\subfigure[]{
\includegraphics[width=7.5cm,height=5cm, angle=0]{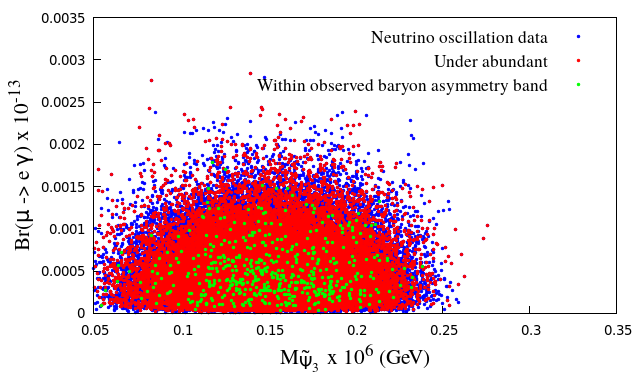}} \\
\subfigure[]{
\includegraphics[width=7.5cm,height=5cm, angle=0]{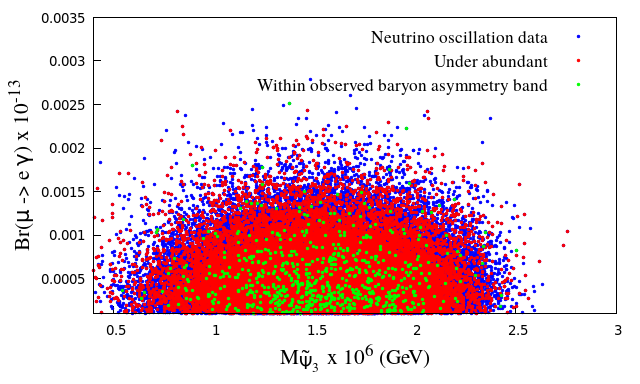}}  
\subfigure[]{
\includegraphics[width=7.5cm,height=5cm, angle=0]{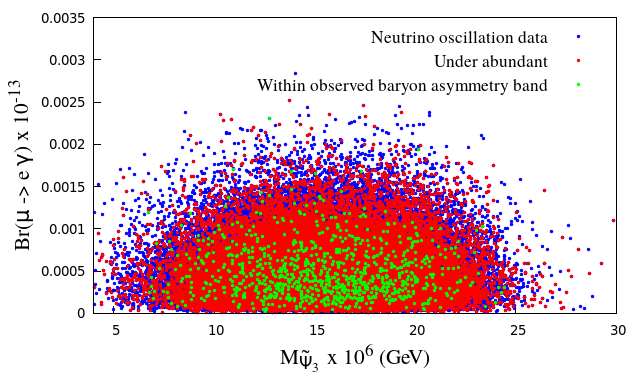}}}
\caption{Variation of BR($\mu \to e \gamma$) with $M_{\tilde{\Psi_3}}$ for (a) $M_{\tilde{\Psi_1}} < 10$ TeV, (upper left) (b) 10 TeV $< M_{\tilde{\Psi_1}} <$ 100 TeV, (upper right) (c) 100 TeV $< M_{\tilde{\Psi_1}} <$ 1000 TeV, (lower left) (d) $M_{\tilde{\Psi_1}} >$ 1000 TeV (lower right). Color coding is expressed in legends.}
\label{Br:m3}
\end{figure}

For these mass intervals, we have checked that for all points in the parameter space BR($l_i \to l_j \gamma$) remains well below the experimental bounds mentioned  in Table \ref{lfv:tab}. Since the strongest bound applies on the LFV process $\mu \to e \gamma$, we have depicted the parameter space in the BR($\mu \to e \gamma$) vs. $M_{\tilde{\Psi_1}}~(M_{\tilde{\Psi_3}})$ plane for the four aforementioned mass regime in Fig.\ref{Br:m1} (Fig.\ref{Br:m3}). In Fig.\ref{Br:m1} and Fig.\ref{Br:m3}, the blue points are allowed by the neutrino oscillation data only. The red points satisfy the neutrino oscillation data and the computed baryon asymmetry is smaller than the lower edge of the observed data as mentioned in Eq.(\ref{baryon-limit}). In other words, the red points are termed as under abundant with respect to the baryon asymmetry. Finally the green points are compatible with the neutrino oscillation data and lie within the experimental band of baryon asymmetry. In Fig.\ref{Br:m1} (Fig.\ref{Br:m3}) (a),(b),(c) and (d)  one can see that for $1 ~ {\rm TeV}~ (4 ~ {\rm TeV}) < M_{\tilde{\Psi_1}}~(M_{\tilde{\Psi_3}}) < 16~ {\rm MeV}~(30~ {\rm MeV} ) $, the red coloured under abundant points satisfy ${\rm Br} (\mu \to e \gamma) < 2.5 \times 10^{-16}$, which is well below the experimental bound prescribed by MEG. The parameter space in the branching ratio vs. $M_{\tilde{\Psi_1}}$ ($M_{\tilde{\Psi_3}}$) plane shrinks after imposing all the aforementioned constraints including the observed baryon asymmetry. Thus for the green points the aforementioned upper edge of ${\rm Br} (\mu \to e \gamma)$ goes down to $\sim 1.5 \times 10^{-16}$. These small values of ${\rm Br} (\mu \to e \gamma)$ are attributed to tiny mixings among the active and the heavy neutrinos. The limits on the branching ratios of other LFV processes ($\tau \to e \gamma,~ \tau \to \mu \gamma$) are satisfied simultaneously, since they are weak compare to the limit on Br($\mu \to e \gamma$). The lowest value of $M_{\tilde{\Psi_1}}$ for which the observed baryon asymmetry can be satisfied is $\sim$ 3.2 TeV.

\begin{figure}[htpb!]{\centering
\subfigure[]{
\includegraphics[width=7.5cm,height=5cm, angle=0]{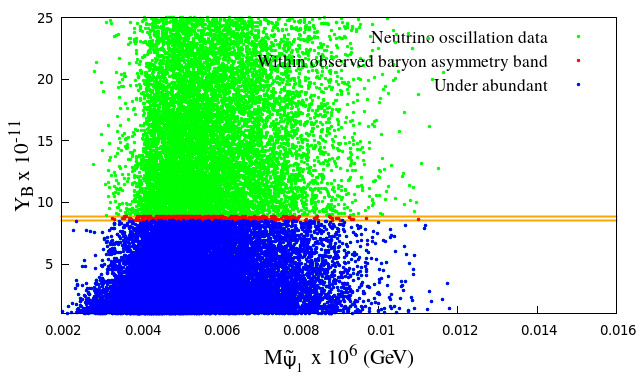}} 
\subfigure[]{
\includegraphics[width=7.5cm,height=5cm, angle=0]{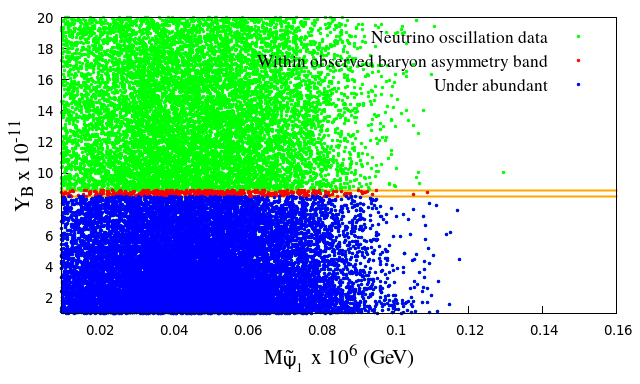}} \\
\subfigure[]{
\includegraphics[width=7.5cm,height=5cm, angle=0]{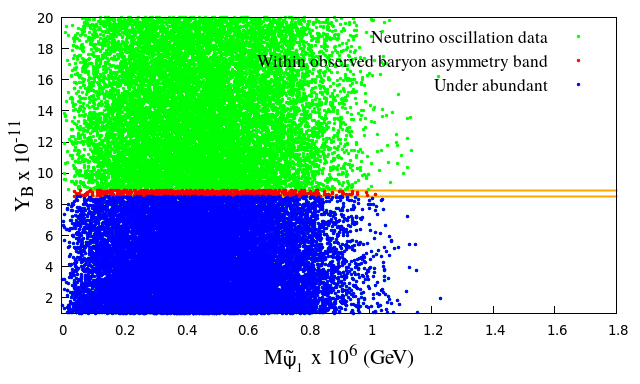}}  
\subfigure[]{
\includegraphics[width=7.5cm,height=5cm, angle=0]{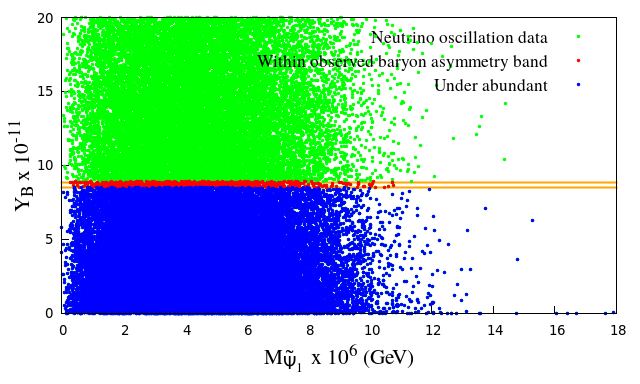}}}
\caption{Variation of $Y_B$ with $M_{\tilde{\Psi_1}}$ for (a) $M_{\tilde{\Psi_1}} < 10$ TeV, (upper left) (b) 10 TeV $< M_{\tilde{\Psi_1}} <$ 100 TeV, (upper right) (c) 100 TeV $< M_{\tilde{\Psi_1}} <$ 1000 TeV, (lower left) (d) $M_{\tilde{\Psi_1}} >$ 1000 TeV (lower right). Color coding is expressed in legends. Orange horizontal lines represent the allowed region of $Y_B$ from the experiments as given in~Eq.(\ref{baryon-limit}).}
\label{YB:m1}
\end{figure}

\begin{figure}[htpb!]{\centering
\subfigure[]{
\includegraphics[width=7.5cm,height=5cm, angle=0]{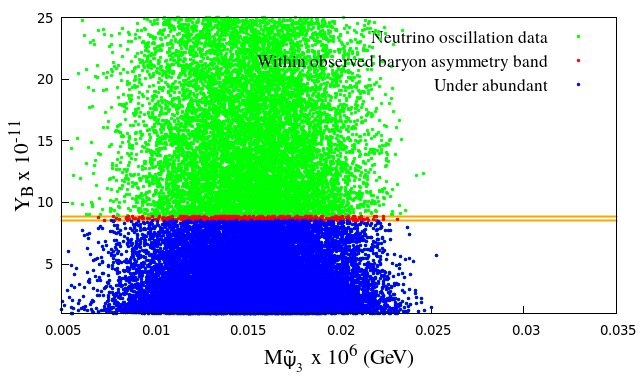}} 
\subfigure[]{
\includegraphics[width=7.5cm,height=5cm, angle=0]{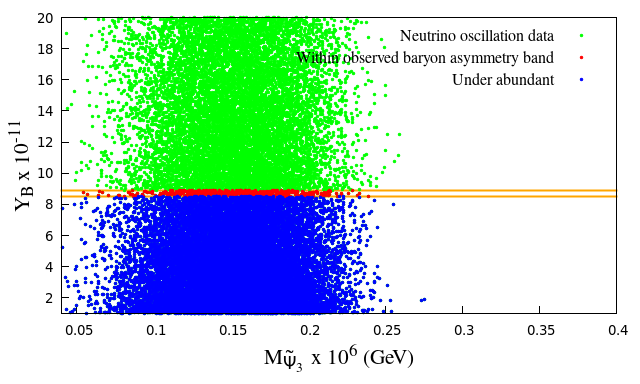}} \\
\subfigure[]{
\includegraphics[width=7.5cm,height=5cm, angle=0]{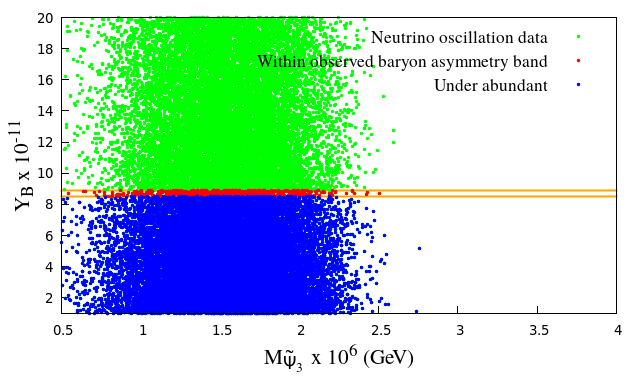}}  
\subfigure[]{
\includegraphics[width=7.5cm,height=5cm, angle=0]{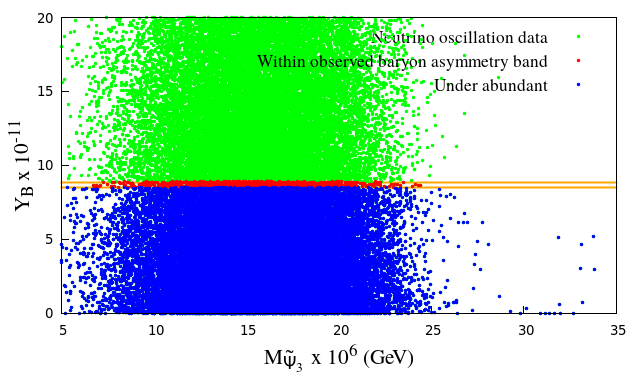}}}
\caption{Variation of $Y_B$ with $M_{\tilde{\Psi_3}}$ for (a) $M_{\tilde{\Psi_1}} < 10$ TeV, (upper left) (b) 10 TeV $< M_{\tilde{\Psi_1}} <$ 100 TeV, (upper right) (c) 100 TeV $< M_{\tilde{\Psi_1}} <$ 1000 TeV, (lower left) (d) $M_{\tilde{\Psi_1}} >$ 1000 TeV (lower right). Color coding is expressed in legends. Orange horizontal lines represent the allowed region of $Y_B$ from the experiments as given in~Eq.(\ref{baryon-limit}).}.
\label{YB:m3}
\end{figure}

Next we show the variation of baryon asymmetry $Y_B$ with $M_{\tilde{\Psi_1}}~(M_{\tilde{\Psi_3}})$ for all mass ranges in Fig.\ref{YB:m1} (Fig.\ref{YB:m3}). The points within the red band in Fig.\ref{YB:m1} and Fig.\ref{YB:m3} are compatible with the neutrino oscillation data, LFV constraints and observed baryon asymmetry. For the blue points, adequate baryon asymmetry cannot be generated due to the interplay of the model parameters. Thus the blue points (under abundant) and the  red points are allowed by the baryon asymmetry data. Whereas the baryon asymmetry for the green points are over abundant and thus are ruled out for this scenario. Thus for a wide mass range, {\em i.e.} [$< $ 10 TeV : $>$ 100 TeV], large part of the parameter space are allowed by the neutrino oscillation data, LFV constraints and observed baryon asymmetry all together.

Before concluding this section, we would like to make a comment on the impact of a non-zero $M_R$ on the results. Here we have checked that following the hierarchy $\mu , M_R << M_D << M_S$, with a non-zero but tiny $M_R$, all the numerical results hardly show any deviation with respect to what is obtained by setting $M_R = 0$. Our initial assumption of taking a vanishing $M_R$ (aimed towards simplifying the analysis) is thus justified.

\section{Conclusion and future outlook}
\label{conclusion}
In this framework, the SM is augmented with two RH neutrinos and two singlet fermions for generating neutrino mass and mixing through inverse see-saw mechanism. This scenario has been popularly termed as (2,2) ISS model in literature. Considering the normal hierarchy among the active neutrino masses and the most general structure of the neutrino mass matrix $M_\nu$ (with complex entries), one can tune all the model parameters in such a way that the multi-dimensional parameter space is compatible with neutrino oscillation data.  As $M_\nu$ is a $7 \times 7 $ complex symmetric matrix, one can diagonalise $M_\nu$ and compute the linear combination of three active neutrinos, two RH neutrinos and two singlet fermions to have seven mass eigenstates. Among these seven mass eigenstates, three corresponds to light active neutrinos and the rest of the four ($\tilde{\Psi_1}, \tilde{\Psi_2}, \tilde{\Psi_3}, \tilde{\Psi_4}$) are heavy. Here one must notice that two of the four mass eigenstates are almost mass degenerate, {\em i.e.} $M_{\tilde{\Psi_1}} \simeq M_{\tilde{\Psi_2}}$ and $M_{\tilde{\Psi_3}} \simeq M_{\tilde{\Psi_4}}$. Next we check if the parameter space compatible with neutrino oscillation data also complies with the experimental bounds coming from the LFV decays. Owing to small mixing among the active and heavy neutrinos, the strongest LFV constraint coming from BR($\mu \to e \gamma$) is satisfied throughout the entire parameter space. Constraints on the branching ratios of the other LFV decays like $\tau \to e \gamma$, $\tau \to \mu \gamma$ are also satisfied. 

Next we compute the baryon asymmetry for each and every point allowed by the neutrino oscillation data and LFV constraints. Here the baryon asymmetry is generated via resonant leptogenesis, where the $CP$-asymmetries corresponding to the lightest mass degenerate pair enter into the Boltzmann equations. For the analysis let us divide $M_{\tilde{\Psi_1}} (\simeq M_{\tilde{\Psi_2}})$ in the following ranges : (a) $M_{\tilde{\Psi_1}} < 10$ TeV, (b) 10 TeV $< M_{\tilde{\Psi_1}} <$ 100 TeV, (c) 100 TeV $< M_{\tilde{\Psi_1}} <$ 1000 TeV, (d) $M_{\tilde{\Psi_1}} >$ 1000 TeV. For these four mass regimes the total parameter space can be splitted into three parts which satisfy : (i) neutrino oscillation data, (ii) neutrino oscillation data and LFV constraints with insufficient baryon asymmetry (under abundant), (iii) neutrino oscillation data, LFV constraints with correct baryon asymmetry (within the observed band). For points belonging to the third part of the parameter space, Br($\mu \to e \gamma$) can be as low as $\sim 1.5 \times 10^{-16}$ due to small mixing between the active and heavy neutrinos. The lowest value of $M_{\tilde{\Psi_1}}$ for which the observed baryon asymmetry can be satisfied is $\sim$ 3.2 TeV. 

Thus to generate adequate baryon asymmetry in this framework, one needs atleast TeV scale heavy neutrinos which could be interesting to look for in the future collider experiments. Another very exciting probe for these models with TeV scale neutrinos might be the observation of neutrino less double beta decay process. The non-observation of the aforementioned signal can put severe constraint on such models with TeV scale neutrinos. Analysis of this particular scenario in light of neurtrino less double beta decay experiment and subsequent collider analysis of relevant signals involving TeV scale neutrinos warrant a separate study in future \cite{collider}.
\section{Acknowledgements}
The authors thank Dr. Joydeep Chakrabortty for a careful reading of the manuscript. IC thanks Dr. Nabarun Chakrabarty for fruitful discussions. IC also acknowledges support from DST, India, under grant number IFA18-PH214 (INSPIRE Faculty Award). HR is supported by the Science and Engineering Research Board,
Government of India, under the agreement SERB/PHY/2016348 (Early Career Research
Award). TS acknowledges the support from the Dr. D. S. Kothari Postdoctoral scheme No. PH/20-21/0163. Finally, IC and TS are thankful to Indian Institute
of Technology Guwahati for organising WHEPP (Workshop on High
Energy Physics Phenomenology) XVI where this work was
initiated.
\appendix
\section{Matrix elements}
\label{app:A}
Let us write down $m_\nu ^{-1}$ as:
\bea
\begin{pmatrix}
a_{1,1}^R + I a_{1,1}^I & ~~a_{1,2}^R + I a_{1,2}^I &~~ a_{1,3}^R + I a_{1,3}^I \\
a_{2,1}^R + I a_{2,1}^I &~~ a_{2,2}^R + I a_{2,2}^I &~~ a_{2,3}^R + I a_{2,3}^I \\
a_{3,1}^R + I a_{3,1}^I &~~ a_{3,2}^R + I a_{3,2}^I &~~ a_{3,3}^R + I a_{3,3}^I 
\end{pmatrix}
\eea
Elements of $M_{\rm mid}$ (with $M_R = 0$) can be written in terms of real and imaginary parts as ,
\bea
{M_{\rm mid}}_{(1,1)}^R &=& 
-2 a_{1,1}^I {M_D}_{1,1}^I {M_D}_{1,1}^R + a_{1,1}^R \left(({M_D}_{1,1}^R)^2-({M_D}_{1,1}^I)^2\right)-2 a_{1,2}^I {M_D}_{1,1}^I {M_D}_{2,1}^R \nonumber \\
&& -2 a_{1,2}^I {M_D}_{1,1}^R {M_D}_{2,1}^I -2 a_{1,2}^R {M_D}_{1,1}^I
{M_D}_{2,1}^I+2 a_{1,2}^R {M_D}_{1,1}^R {M_D}_{2,1}^R \nonumber \\
&& -2 {M_D}_{3,1}^R (a_{1,3}^I {M_D}_{1,1}^I-a_{1,3}^R {M_D}_{1,1}^R+a_{2,3}^I {M_D}_{2,1}^I
  - a_{2,3}^R {M_D}_{2,1}^R + a_{3,3}^I {M_D}_{3,1}^I) \nonumber \\
  && -2 a_{1,3}^I
  {M_D}_{1,1}^R {M_D}_{3,1}^I-2 a_{1,3}^R {M_D}_{1,1}^I {M_D}_{3,1}^I-2 a_{2,2}^I {M_D}_{2,1}^I {M_D}_{2,1}^R-a_{2,2}^R ({M_D}_{2,1}^I)^2 \nonumber \\
&& +a_{2,2}^R ({M_D}_{2,1}^R)^2-2 a_{2,3}^I {M_D}_{2,1}^R {M_D}_{3,1}^I-2
   a_{2,3}^R {M_D}_{2,1}^I {M_D}_{3,1}^I-a_{3,3}^R ({M_D}_{3,1}^I)^2 \nonumber\\ && +a_{3,3}^R ({M_D}_{3,1}^R)^2 \, 
   \eea
\bea
{M_{\rm mid}}_{(1,1)}^I &=&   
a_{1,1}^I \left(({M_D}_{1,1}^R)^2-({M_D}_{1,1}^I)^2\right)+2 a_{1,1}^R {M_D}_{1,1}^I {M_D}_{1,1}^R-2 a_{1,2}^I {M_D}_{1,1}^I {M_D}_{2,1}^I \nonumber \\
&& +2 a_{1,2}^I {M_D}_{1,1}^R {M_D}_{2,1}^R+2 a_{1,2}^R {M_D}_{1,1}^I
   {M_D}_{2,1}^R+2 a_{1,2}^R {M_D}_{1,1}^R {M_D}_{2,1}^I \nonumber \\
   && +2 {M_D}_{3,1}^R (a_{1,3}^I {M_D}_{1,1}^R+a_{1,3}^R {M_D}_{1,1}^I+a_{2,3}^I {M_D}_{2,1}^R+a_{2,3}^R {M_D}_{2,1}^I+a_{3,3}^R {M_D}_{3,1}^I)\nonumber \\
   &&-2 a_{1,3}^I
   {M_D}_{1,1}^I {M_D}_{3,1}^I+2 a_{1,3}^R {M_D}_{1,1}^R {M_D}_{3,1}^I-a_{2,2}^I ({M_D}_{2,1}^I)^2+a_{2,2}^I ({M_D}_{2,1}^R)^2 \nonumber \\
   && +2 a_{2,2}^R {M_D}_{2,1}^I {M_D}_{2,1}^R-2 a_{2,3}^I {M_D}_{2,1}^I {M_D}_{3,1}^I+2
   a_{2,3}^R {M_D}_{2,1}^R {M_D}_{3,1}^I-a_{3,3}^I ({M_D}_{3,1}^I)^2 \nonumber \\
   &&+a_{3,3}^I ({M_D}_{3,1}^R)^2   \,
   \eea
   \bea
{M_{\rm mid}}_{(1,2)}^R &=&       
   -a_{1,1}^I ({M_D}_{1,1}^I {M_D}_{1,2}^R+{M_D}_{1,1}^R {M_D}_{1,2}^I)+a_{1,1}^R ({M_D}_{1,1}^R {M_D}_{1,2}^R-{M_D}_{1,1}^I
   {M_D}_{1,2}^I) \nonumber \\
   && -a_{1,2}^I {M_D}_{1,1}^I {M_D}_{2,2}^R-a_{1,2}^I {M_D}_{1,1}^R {M_D}_{2,2}^I-a_{1,2}^I {M_D}_{1,2}^I
   {M_D}_{2,1}^R-a_{1,2}^I {M_D}_{1,2}^R {M_D}_{2,1}^I \nonumber \\
   && -a_{1,2}^R {M_D}_{1,1}^I {M_D}_{2,2}^I+a_{1,2}^R {M_D}_{1,1}^R
   {M_D}_{2,2}^R-a_{1,2}^R {M_D}_{1,2}^I {M_D}_{2,1}^I+a_{1,2}^R {M_D}_{1,2}^R {M_D}_{2,1}^R \nonumber \\
   && +{M_D}_{3,2}^R (-a_{1,3}^I
   {M_D}_{1,1}^I+a_{1,3}^R {M_D}_{1,1}^R-a_{2,3}^I {M_D}_{2,1}^I+a_{2,3}^R {M_D}_{2,1}^R-a_{3,3}^I {M_D}_{3,1}^I+a_{3,3}^R
   {M_D}_{3,1}^R) \nonumber \\
  && -a_{1,3}^I {M_D}_{1,1}^R {M_D}_{3,2}^I-a_{1,3}^I {M_D}_{1,2}^I {M_D}_{3,1}^R-a_{1,3}^I {M_D}_{1,2}^R
   {M_D}_{3,1}^I-a_{1,3}^R {M_D}_{1,1}^I {M_D}_{3,2}^I \nonumber \\
   && -a_{1,3}^R {M_D}_{1,2}^I {M_D}_{3,1}^I+a_{1,3}^R {M_D}_{1,2}^R
   {M_D}_{3,1}^R-a_{2,2}^I {M_D}_{2,1}^I {M_D}_{2,2}^R-a_{2,2}^I {M_D}_{2,1}^R {M_D}_{2,2}^I \nonumber \\
   && -a_{2,2}^R {M_D}_{2,1}^I
   {M_D}_{2,2}^I+a_{2,2}^R {M_D}_{2,1}^R {M_D}_{2,2}^R-a_{2,3}^I {M_D}_{2,1}^R {M_D}_{3,2}^I-a_{2,3}^I {M_D}_{2,2}^I
   {M_D}_{3,1}^R \nonumber \\
   && -a_{2,3}^I {M_D}_{2,2}^R {M_D}_{3,1}^I-a_{2,3}^R {M_D}_{2,1}^I {M_D}_{3,2}^I-a_{2,3}^R {M_D}_{2,2}^I
   {M_D}_{3,1}^I+a_{2,3}^R {M_D}_{2,2}^R {M_D}_{3,1}^R \nonumber \\
   && -a_{3,3}^I {M_D}_{3,1}^R {M_D}_{3,2}^I-a_{3,3}^R {M_D}_{3,1}^I
   {M_D}_{3,2}^I \, 
   \eea
   \bea
   {M_{\rm mid}}_{(1,2)}^I &=&   -a_{1,1}^I {M_D}_{1,1}^I {M_D}_{1,2}^I+a_{1,1}^I {M_D}_{1,1}^R {M_D}_{1,2}^R+a_{1,1}^R {M_D}_{1,1}^I {M_D}_{1,2}^R+a_{1,1}^R
   {M_D}_{1,1}^R {M_D}_{1,2}^I \nonumber \\
   && -a_{1,2}^I {M_D}_{1,1}^I {M_D}_{2,2}^I+a_{1,2}^I {M_D}_{1,1}^R {M_D}_{2,2}^R-a_{1,2}^I {M_D}_{1,2}^I
   {M_D}_{2,1}^I+a_{1,2}^I {M_D}_{1,2}^R {M_D}_{2,1}^R \nonumber \\
   &&+a_{1,2}^R {M_D}_{1,1}^I {M_D}_{2,2}^R+a_{1,2}^R {M_D}_{1,1}^R
   {M_D}_{2,2}^I 
    +a_{1,2}^R {M_D}_{1,2}^I {M_D}_{2,1}^R+a_{1,2}^R {M_D}_{1,2}^R {M_D}_{2,1}^I 
  \nonumber \\
  && +{M_D}_{3,2}^R (a_{1,3}^I 
   {M_D}_{1,1}^R 
   +a_{1,3}^R {M_D}_{1,1}^I+a_{2,3}^I {M_D}_{2,1}^R+a_{2,3}^R {M_D}_{2,1}^I+a_{3,3}^I {M_D}_{3,1}^R+a_{3,3}^R
   {M_D}_{3,1}^I) \nonumber \\
   &&-a_{1,3}^I {M_D}_{1,1}^I {M_D}_{3,2}^I-a_{1,3}^I {M_D}_{1,2}^I {M_D}_{3,1}^I+a_{1,3}^I {M_D}_{1,2}^R
   {M_D}_{3,1}^R+a_{1,3}^R {M_D}_{1,1}^R {M_D}_{3,2}^I \nonumber \\
   && +a_{1,3}^R {M_D}_{1,2}^I {M_D}_{3,1}^R 
    +a_{1,3}^R {M_D}_{1,2}^R
   {M_D}_{3,1}^I-a_{2,2}^I {M_D}_{2,1}^I {M_D}_{2,2}^I +a_{2,2}^I {M_D}_{2,1}^R {M_D}_{2,2}^R \nonumber \\
   &&+a_{2,2}^R {M_D}_{2,1}^I
   {M_D}_{2,2}^R+a_{2,2}^R {M_D}_{2,1}^R {M_D}_{2,2}^I-a_{2,3}^I {M_D}_{2,1}^I {M_D}_{3,2}^I-a_{2,3}^I {M_D}_{2,2}^I
   {M_D}_{3,1}^I \nonumber \\
   &&+a_{2,3}^I {M_D}_{2,2}^R {M_D}_{3,1}^R +a_{2,3}^R {M_D}_{2,1}^R {M_D}_{3,2}^I+a_{2,3}^R {M_D}_{2,2}^I
   {M_D}_{3,1}^R+a_{2,3}^R {M_D}_{2,2}^R {M_D}_{3,1}^I \nonumber \\
   &&-a_{3,3}^I {M_D}_{3,1}^I {M_D}_{3,2}^I+a_{3,3}^R {M_D}_{3,1}^R {M_D}_{3,2}^I \, 
   \eea
   \bea
   {M_{\rm mid}}_{(2,2)}^R &=&     
-2 a_{1,1}^I {M_D}_{1,2}^I {M_D}_{1,2}^R+a_{1,1}^R \left(({M_D}_{1,2}^R)^2-({M_D}_{1,2}^I)^2\right)-2 a_{1,2}^I {M_D}_{1,2}^I {M_D}_{2,2}^R \nonumber \\
&& -2
   a_{1,2}^I {M_D}_{1,2}^R {M_D}_{2,2}^I-2 a_{1,2}^R {M_D}_{1,2}^I {M_D}_{2,2}^I+2 a_{1,2}^R {M_D}_{1,2}^R {M_D}_{2,2}^R \nonumber \\
   && -2 {M_D}_{3,2}^R
   (a_{1,3}^I {M_D}_{1,2}^I-a_{1,3}^R {M_D}_{1,2}^R+a_{2,3}^I {M_D}_{2,2}^I-a_{2,3}^R {M_D}_{2,2}^R+a_{3,3}^I {M_D}_{3,2}^I) \nonumber \\
   &&-2
   a_{1,3}^I {M_D}_{1,2}^R {M_D}_{3,2}^I-2 a_{1,3}^R {M_D}_{1,2}^I {M_D}_{3,2}^I-2 a_{2,2}^I {M_D}_{2,2}^I {M_D}_{2,2}^R-a_{2,2}^R
   ({M_D}_{2,2}^I)^2 \nonumber \\
   && +a_{2,2}^R ({M_D}_{2,2}^R)^2-2 a_{2,3}^I {M_D}_{2,2}^R {M_D}_{3,2}^I-2 a_{2,3}^R {M_D}_{2,2}^I {M_D}_{3,2}^I-a_{3,3}^R
   ({M_D}_{3,2}^I)^2 \nonumber \\
   && +a_{3,3}^R ({M_D}_{3,2}^R)^2 \, 
   \eea   
   \bea
  {M_{\rm mid}}_{(2,2)}^I &=&  a_{1,1}^I ({M_D}_{1,2}^R)^2-({M_D}_{1,2}^I)^2)+2 a_{1,1}^R {M_D}_{1,2}^I {M_D}_{1,2}^R-2 a_{1,2}^I {M_D}_{1,2}^I {M_D}_{2,2}^I+2 a_{1,2}^I {M_D}_{1,2}^R {M_D}_{2,2}^R\nonumber \\
  && +2 a_{1,2}^R {M_D}_{1,2}^I  {M_D}_{2,2}^R
  +2 a_{1,2}^R {M_D}_{1,2}^R {M_D}_{2,2}^I
  +2 {M_D}_{3,2}^R (a_{1,3}^I {M_D}_{1,2}^R
  +a_{1,3}^R {M_D}_{1,2}^I
  +a_{2,3}^I {M_D}_{2,2}^R \nonumber \\
  && +a_{2,3}^R {M_D}_{1,2}^I+a_{3,3}^R {M_D}_{3,2}^I)-2 a_{1,3}^I
   {M_D}_{1,2}^I {M_D}_{3,2}^I
   +2 a_{1,3}^R {M_D}_{1,2}^R {M_D}_{3,2}^I
   -a_{2,2}^I ({M_D}_{2,2}^I)^2 \nonumber \\
 &&  +a_{2,2}^I ({M_D}_{2,2}^R)^2+2 a_{2,2}^R {M_D}_{2,2}^I {M_D}_{2,2}^R-2 a_{2,3}^I {M_D}_{2,2}^I {M_D}_{3,2}^I+2
   a_{2,3}^R {M_D}_{2,2}^R {M_D}_{3,2}^I \nonumber \\
   &&-a_{3,3}^I ({M_D}_{3,2}^I)^2+a_{3,3}^I ({M_D}_{3,2}^R)^2
   \eea

Now real and imaginary parts of the elements of $\mu$ are, 

\bea
&& \mu_{1,1}^R = \frac{A_{1,1}^R}{B_{1,1}^R} , ~~~ \mu_{1,1}^I = \frac{A_{1,1}^I}{B_{1,1}^I} \nonumber \\
&& \mu_{1,2}^R = \frac{A_{1,2}^R}{B_{1,2}^R} , ~~~ \mu_{1,2}^I = \frac{A_{1,2}^I}{B_{1,2}^I}  \nonumber \\
&& \mu_{2,2}^R = \frac{A_{2,2}^R}{B_{2,2}^R} , ~~~ \mu_{2,2}^I = \frac{A_{2,2}^I}{B_{2,2}^I}
\eea

where,

\bea
A_{1,1}^R &=& 2 {M_S}_{1,2}^R ({M_S}_{1,2}^I ({M_{\rm mid}}_{(2,2)}^I (({M_{\rm mid}}_{(1,1)}^I)^2+({M_{\rm mid}}_{(1,1)}^R)^2)+{M_{\rm mid}}_{(1,1)}^I ({M_{\rm mid}}_{(1,2)}^R)^2) \nonumber \\
&& +{M_{\rm mid}}_{(1,2)}^R {M_S}_{1,1}^R ({M_{\rm mid}}_{(1,1)}^I
   {M_{\rm mid}}_{(2,2)}^I-{M_{\rm mid}}_{(1,1)}^R {M_{\rm mid}}_{(2,2)}^R+({M_{\rm mid}}_{(1,2)}^R)^2) \nonumber \\
   &&-{M_{\rm mid}}_{(1,2)}^R {M_S}_{1,1}^I ({M_{\rm mid}}_{(1,1)}^I {M_{\rm mid}}_{(2,2)}^R+{M_{\rm mid}}_{(1,1)}^R {M_{\rm mid}}_{(2,2)}^I))+({M_S}_{1,2}^R)^2 ({M_{\rm mid}}_{(2,2)}^R
   (({M_{\rm mid}}_{(1,1)}^I)^2 \nonumber \\
   &&+({M_{\rm mid}}_{(1,1)}^R)^2)-{M_{\rm mid}}_{(1,1)}^R ({M_{\rm mid}}_{(1,2)}^R)^2) -({M_{\rm mid}}_{(1,1)}^I)^2 {M_{\rm mid}}_{(2,2)}^R ({M_S}_{1,2}^I)^2 \nonumber \\
   &&-({M_{\rm mid}}_{(1,2)}^I)^2 (2 {M_{\rm mid}}_{(1,1)}^I {M_S}_{1,2}^I
   {M_S}_{1,2}^R +{M_{\rm mid}}_{(1,1)}^R ({M_S}_{1,2}^I)^2-{M_{\rm mid}}_{(1,1)}^R ({M_S}_{1,2}^R)^2 \nonumber \\
   &&+2 {M_{\rm mid}}_{(1,2)}^R {M_S}_{1,1}^I {M_S}_{1,2}^I-2 {M_{\rm mid}}_{(1,2)}^R {M_S}_{1,1}^R {M_S}_{1,2}^R +2 {M_{\rm mid}}_{(2,2)}^I {M_S}_{1,1}^I
   {M_S}_{1,1}^R+{M_{\rm mid}}_{(2,2)}^R ({M_S}_{1,1}^I \nonumber \\
   && -{M_S}_{1,1}^R) ({M_S}_{1,1}^I+{M_S}_{1,1}^R))+2 {M_{\rm mid}}_{(1,2)}^I ({M_{\rm mid}}_{(1,2)}^R ({M_{\rm mid}}_{(1,1)}^I ({M_S}_{1,2}^I-{M_S}_{1,2}^R) ({M_S}_{1,2}^I+{M_S}_{1,2}^R) \nonumber \\
   && -2 {M_{\rm mid}}_{(1,1)}^R {M_S}_{1,2}^I {M_S}_{1,2}^R+{M_{\rm mid}}_{(2,2)}^I ({M_S}_{1,1}^I-{M_S}_{1,1}^R) ({M_S}_{1,1}^I+{M_S}_{1,1}^R)-2 {M_{\rm mid}}_{(2,2)}^R {M_S}_{1,1}^I {M_S}_{1,1}^R) \nonumber \\
   &&+{M_S}_{1,2}^I (-{M_{\rm mid}}_{(1,1)}^I {M_{\rm mid}}_{(2,2)}^I
   {M_S}_{1,1}^R+{M_{\rm mid}}_{(1,1)}^I {M_{\rm mid}}_{(2,2)}^R {M_S}_{1,1}^I+{M_{\rm mid}}_{(1,1)}^R {M_{\rm mid}}_{(2,2)}^I {M_S}_{1,1}^I \nonumber \\
   &&+{M_{\rm mid}}_{(1,1)}^R {M_{\rm mid}}_{(2,2)}^R {M_S}_{1,1}^R)-{M_S}_{1,2}^R ({M_{\rm mid}}_{(1,1)}^I {M_{\rm mid}}_{(2,2)}^I
   {M_S}_{1,1}^I+{M_{\rm mid}}_{(1,1)}^I {M_{\rm mid}}_{(2,2)}^R {M_S}_{1,1}^R \nonumber \\
   && +{M_{\rm mid}}_{(1,1)}^R {M_{\rm mid}}_{(2,2)}^I {M_S}_{1,1}^R-{M_{\rm mid}}_{(1,1)}^R {M_{\rm mid}}_{(2,2)}^R {M_S}_{1,1}^I)+({M_{\rm mid}}_{(1,2)}^R)^2 ({M_S}_{1,1}^I {M_S}_{1,2}^R+{M_S}_{1,1}^R
   {M_S}_{1,2}^I)) \nonumber \\
   &&-2 {M_{\rm mid}}_{(1,1)}^I {M_{\rm mid}}_{(1,2)}^R {M_{\rm mid}}_{(2,2)}^I {M_S}_{1,1}^I {M_S}_{1,2}^I-2 {M_{\rm mid}}_{(1,1)}^I {M_{\rm mid}}_{(1,2)}^R {M_{\rm mid}}_{(2,2)}^R {M_S}_{1,1}^R {M_S}_{1,2}^I \nonumber \\
   && +2 {M_{\rm mid}}_{(1,1)}^I ({M_{\rm mid}}_{(2,2)}^I)^2
   {M_S}_{1,1}^I {M_S}_{1,1}^R+2 {M_{\rm mid}}_{(1,1)}^I ({M_{\rm mid}}_{(2,2)}^R)^2 {M_S}_{1,1}^I {M_S}_{1,1}^R \nonumber \\
   &&-({M_{\rm mid}}_{(1,1)}^R)^2 {M_{\rm mid}}_{(2,2)}^R ({M_S}_{1,2}^I)^2+{M_{\rm mid}}_{(1,1)}^R ({M_{\rm mid}}_{(1,2)}^R)^2 ({M_S}_{1,2}^I)^2 \nonumber \\
   &&-2 {M_{\rm mid}}_{(1,1)}^R
   {M_{\rm mid}}_{(1,2)}^R {M_{\rm mid}}_{(2,2)}^I {M_S}_{1,1}^R {M_S}_{1,2}^I +2 {M_{\rm mid}}_{(1,1)}^R {M_{\rm mid}}_{(1,2)}^R {M_{\rm mid}}_{(2,2)}^R {M_S}_{1,1}^I {M_S}_{1,2}^I \nonumber \\
   &&-{M_{\rm mid}}_{(1,1)}^R ({M_{\rm mid}}_{(2,2)}^I)^2 ({M_S}_{1,1}^I)^2+{M_{\rm mid}}_{(1,1)}^R ({M_{\rm mid}}_{(2,2)}^I)^2
   ({M_S}_{1,1}^R)^2-{M_{\rm mid}}_{(1,1)}^R ({M_{\rm mid}}_{(2,2)}^R)^2 ({M_S}_{1,1}^I)^2 \nonumber \\
   && +{M_{\rm mid}}_{(1,1)}^R ({M_{\rm mid}}_{(2,2)}^R)^2 ({M_S}_{1,1}^R)^2+2 ({M_{\rm mid}}_{(1,2)}^I)^3 ({M_S}_{1,1}^I {M_S}_{1,2}^R+{M_S}_{1,1}^R {M_S}_{1,2}^I) \nonumber \\
   &&-2 ({M_{\rm mid}}_{(1,2)}^R)^3
   {M_S}_{1,1}^I {M_S}_{1,2}^I+2 ({M_{\rm mid}}_{(1,2)}^R)^2 {M_{\rm mid}}_{(2,2)}^I {M_S}_{1,1}^I {M_S}_{1,1}^R+({M_{\rm mid}}_{(1,2)}^R)^2 {M_{\rm mid}}_{(2,2)}^R ({M_S}_{1,1}^I)^2 \nonumber \\
   &&-({M_{\rm mid}}_{(1,2)}^R)^2 {M_{\rm mid}}_{(2,2)}^R
   ({M_S}_{1,1}^R)^2
   \eea
   \bea
   B_{1,1}^R &=& (({M_{\rm mid}}_{(1,1)}^I)^2+({M_{\rm mid}}_{(1,1)}^R)^2) (({M_{\rm mid}}_{(2,2)}^I)^2+({M_{\rm mid}}_{(2,2)}^R)^2)+2 ({M_{\rm mid}}_{(1,2)}^I)^2 (-{M_{\rm mid}}_{(1,1)}^I {M_{\rm mid}}_{(2,2)}^I \nonumber \\
   && +{M_{\rm mid}}_{(1,1)}^R
   {M_{\rm mid}}_{(2,2)}^R+({M_{\rm mid}}_{(1,2)}^R)^2)-4 {M_{\rm mid}}_{(1,2)}^I {M_{\rm mid}}_{(1,2)}^R ({M_{\rm mid}}_{(1,1)}^I {M_{\rm mid}}_{(2,2)}^R \nonumber \\
   && +{M_{\rm mid}}_{(1,1)}^R {M_{\rm mid}}_{(2,2)}^I)+2 ({M_{\rm mid}}_{(1,2)}^R)^2 ({M_{\rm mid}}_{(1,1)}^I {M_{\rm mid}}_{(2,2)}^I-{M_{\rm mid}}_{(1,1)}^R
   {M_{\rm mid}}_{(2,2)}^R)+({M_{\rm mid}}_{(1,2)}^I)^4 \nonumber \\
   &&+({M_{\rm mid}}_{(1,2)}^R)^4 \, 
   \eea
\bea
   A_{1,1}^I &=& -{M_{\rm mid}}_{(2,2)}^I ({M_S}_{1,2}^R)^2 (({M_{\rm mid}}_{(1,1)}^I)^2+({M_{\rm mid}}_{(1,1)}^R)^2)+2 {M_{\rm mid}}_{(2,2)}^R {M_S}_{1,2}^I {M_S}_{1,2}^R (({M_{\rm mid}}_{(1,1)}^I)^2 \nonumber \\
   &&+({M_{\rm mid}}_{(1,1)}^R)^2)+({M_{\rm mid}}_{(1,1)}^I)^2 {M_{\rm mid}}_{(2,2)}^I
   ({M_S}_{1,2}^I)^2+({M_{\rm mid}}_{(1,2)}^I)^2 (-{M_{\rm mid}}_{(1,1)}^I ({M_S}_{1,2}^I)^2 \nonumber \\
   && +{M_{\rm mid}}_{(1,1)}^I ({M_S}_{1,2}^R)^2+2 {M_{\rm mid}}_{(1,1)}^R {M_S}_{1,2}^I {M_S}_{1,2}^R+2 {M_{\rm mid}}_{(1,2)}^R {M_S}_{1,1}^I {M_S}_{1,2}^R+2 {M_{\rm mid}}_{(1,2)}^R
   {M_S}_{1,1}^R {M_S}_{1,2}^I\nonumber \\
   &&+{M_{\rm mid}}_{(2,2)}^I (({M_S}_{1,1}^R)^2-({M_S}_{1,1}^I)^2)+2 {M_{\rm mid}}_{(2,2)}^R {M_S}_{1,1}^I {M_S}_{1,1}^R)\nonumber \\
   &&-2 {M_{\rm mid}}_{(1,2)}^I ({M_{\rm mid}}_{(1,2)}^R (2 {M_{\rm mid}}_{(1,1)}^I {M_S}_{1,2}^I
   {M_S}_{1,2}^R +{M_{\rm mid}}_{(1,1)}^R ({M_S}_{1,2}^I \nonumber \\
   &&-{M_S}_{1,2}^R) ({M_S}_{1,2}^I+{M_S}_{1,2}^R)+2 {M_{\rm mid}}_{(2,2)}^I {M_S}_{1,1}^I {M_S}_{1,1}^R+{M_{\rm mid}}_{(2,2)}^R ({M_S}_{1,1}^I-{M_S}_{1,1}^R)
   ({M_S}_{1,1}^I+{M_S}_{1,1}^R)) \nonumber \\
   &&+{M_S}_{1,2}^I ({M_{\rm mid}}_{(1,1)}^I {M_{\rm mid}}_{(2,2)}^I {M_S}_{1,1}^I +{M_{\rm mid}}_{(1,1)}^I {M_{\rm mid}}_{(2,2)}^R {M_S}_{1,1}^R+{M_{\rm mid}}_{(1,1)}^R {M_{\rm mid}}_{(2,2)}^I {M_S}_{1,1}^R \nonumber \\
   &&-{M_{\rm mid}}_{(1,1)}^R {M_{\rm mid}}_{(2,2)}^R
   {M_S}_{1,1}^I)+{M_S}_{1,2}^R (-{M_{\rm mid}}_{(1,1)}^I {M_{\rm mid}}_{(2,2)}^I {M_S}_{1,1}^R \nonumber \\
   &&+{M_{\rm mid}}_{(1,1)}^I {M_{\rm mid}}_{(2,2)}^R {M_S}_{1,1}^I+{M_{\rm mid}}_{(1,1)}^R {M_{\rm mid}}_{(2,2)}^I {M_S}_{1,1}^I+{M_{\rm mid}}_{(1,1)}^R {M_{\rm mid}}_{(2,2)}^R
   {M_S}_{1,1}^R) \nonumber \\
   &&+({M_{\rm mid}}_{(1,2)}^R)^2 ({M_S}_{1,1}^R {M_S}_{1,2}^R-{M_S}_{1,1}^I {M_S}_{1,2}^I))+({M_{\rm mid}}_{(1,2)}^R)^2 ({M_{\rm mid}}_{(1,1)}^I ({M_S}_{1,2}^I-{M_S}_{1,2}^R) ({M_S}_{1,2}^I \nonumber \\
   &&+{M_S}_{1,2}^R)-2 {M_{\rm mid}}_{(1,1)}^R
   {M_S}_{1,2}^I {M_S}_{1,2}^R+{M_{\rm mid}}_{(2,2)}^I ({M_S}_{1,1}^I-{M_S}_{1,1}^R) ({M_S}_{1,1}^I+{M_S}_{1,1}^R)\nonumber \\
   &&-2 {M_{\rm mid}}_{(2,2)}^R {M_S}_{1,1}^I {M_S}_{1,1}^R)-2 {M_{\rm mid}}_{(1,2)}^R {M_S}_{1,2}^I (-{M_{\rm mid}}_{(1,1)}^I {M_{\rm mid}}_{(2,2)}^I
   {M_S}_{1,1}^R+{M_{\rm mid}}_{(1,1)}^I {M_{\rm mid}}_{(2,2)}^R {M_S}_{1,1}^I \nonumber \\
   &&+{M_{\rm mid}}_{(1,1)}^R {M_{\rm mid}}_{(2,2)}^I {M_S}_{1,1}^I+{M_{\rm mid}}_{(1,1)}^R {M_{\rm mid}}_{(2,2)}^R {M_S}_{1,1}^R)+2 {M_{\rm mid}}_{(1,2)}^R {M_S}_{1,2}^R ({M_{\rm mid}}_{(1,1)}^I {M_{\rm mid}}_{(2,2)}^I
   {M_S}_{1,1}^I \nonumber \\
   &&+{M_{\rm mid}}_{(1,1)}^I {M_{\rm mid}}_{(2,2)}^R {M_S}_{1,1}^R+{M_{\rm mid}}_{(1,1)}^R {M_{\rm mid}}_{(2,2)}^I {M_S}_{1,1}^R-{M_{\rm mid}}_{(1,1)}^R {M_{\rm mid}}_{(2,2)}^R {M_S}_{1,1}^I) \nonumber \\
   &&+{M_{\rm mid}}_{(1,1)}^I ({M_{\rm mid}}_{(2,2)}^I)^2 ({M_S}_{1,1}^I)^2-{M_{\rm mid}}_{(1,1)}^I
   ({M_{\rm mid}}_{(2,2)}^I)^2 ({M_S}_{1,1}^R)^2 \nonumber \\
   &&+{M_{\rm mid}}_{(1,1)}^I ({M_{\rm mid}}_{(2,2)}^R)^2 ({M_S}_{1,1}^I)^2-{M_{\rm mid}}_{(1,1)}^I ({M_{\rm mid}}_{(2,2)}^R)^2 ({M_S}_{1,1}^R)^2+({M_{\rm mid}}_{(1,1)}^R)^2 {M_{\rm mid}}_{(2,2)}^I ({M_S}_{1,2}^I)^2 \nonumber \\
   &&+2 {M_{\rm mid}}_{(1,1)}^R ({M_{\rm mid}}_{(2,2)}^I)^2
   {M_S}_{1,1}^I {M_S}_{1,1}^R+2 {M_{\rm mid}}_{(1,1)}^R ({M_{\rm mid}}_{(2,2)}^R)^2 {M_S}_{1,1}^I {M_S}_{1,1}^R \nonumber \\
   &&+2 ({M_{\rm mid}}_{(1,2)}^I)^3 ({M_S}_{1,1}^I {M_S}_{1,2}^I-{M_S}_{1,1}^R {M_S}_{1,2}^R)+2 ({M_{\rm mid}}_{(1,2)}^R)^3 ({M_S}_{1,1}^I
   {M_S}_{1,2}^R+{M_S}_{1,1}^R {M_S}_{1,2}^I) \, \\
   B_{1,1}^I &=& (({M_{\rm mid}}_{(1,1)}^I)^2+({M_{\rm mid}}_{(1,1)}^R)^2) (({M_{\rm mid}}_{(2,2)}^I)^2+({M_{\rm mid}}_{(2,2)}^R)^2)+2 ({M_{\rm mid}}_{(1,2)}^I)^2 (-{M_{\rm mid}}_{(1,1)}^I {M_{\rm mid}}_{(2,2)}^I \nonumber \\
   && +{M_{\rm mid}}_{(1,1)}^R
   {M_{\rm mid}}_{(2,2)}^R+({M_{\rm mid}}_{(1,2)}^R)^2)-4 {M_{\rm mid}}_{(1,2)}^I {M_{\rm mid}}_{(1,2)}^R ({M_{\rm mid}}_{(1,1)}^I {M_{\rm mid}}_{(2,2)}^R \nonumber \\
   &&+{M_{\rm mid}}_{(1,1)}^R {M_{\rm mid}}_{(2,2)}^I)+2 ({M_{\rm mid}}_{(1,2)}^R)^2 ({M_{\rm mid}}_{(1,1)}^I {M_{\rm mid}}_{(2,2)}^I-{M_{\rm mid}}_{(1,1)}^R
   {M_{\rm mid}}_{(2,2)}^R)+({M_{\rm mid}}_{(1,2)}^I)^4 \nonumber \\
   &&+({M_{\rm mid}}_{(1,2)}^R)^4 \, 
   \eea  
\bea
   A_{1,2}^R &=& {M_S}_{2,2}^R (({M_{\rm mid}}_{(1,1)}^I)^2 ({M_{\rm mid}}_{(2,2)}^I {M_S}_{1,2}^I+{M_{\rm mid}}_{(2,2)}^R {M_S}_{1,2}^R)+{M_{\rm mid}}_{(1,1)}^I {M_{\rm mid}}_{(1,2)}^R ({M_{\rm mid}}_{(1,2)}^R {M_S}_{1,2}^I \nonumber \\
   &&+{M_{\rm mid}}_{(2,2)}^I {M_S}_{1,1}^R-{M_{\rm mid}}_{(2,2)}^R
   {M_S}_{1,1}^I)+({M_{\rm mid}}_{(1,2)}^R)^3 {M_S}_{1,1}^R)+({M_{\rm mid}}_{(1,1)}^I)^2 {M_{\rm mid}}_{(2,2)}^I {M_S}_{1,2}^R {M_S}_{2,2}^I \nonumber \\
   && -({M_{\rm mid}}_{(1,1)}^I)^2 {M_{\rm mid}}_{(2,2)}^R {M_S}_{1,2}^I {M_S}_{2,2}^I-({M_{\rm mid}}_{(1,2)}^I)^2 ({M_{\rm mid}}_{(1,1)}^I
   {M_S}_{1,2}^I {M_S}_{2,2}^R+{M_{\rm mid}}_{(1,1)}^I {M_S}_{1,2}^R {M_S}_{2,2}^I \nonumber \\
   &&+{M_{\rm mid}}_{(1,1)}^R {M_S}_{1,2}^I {M_S}_{2,2}^I-{M_{\rm mid}}_{(1,1)}^R {M_S}_{1,2}^R {M_S}_{2,2}^R+{M_{\rm mid}}_{(1,2)}^R {M_S}_{1,1}^I {M_S}_{2,2}^I-{M_{\rm mid}}_{(1,2)}^R
   {M_S}_{1,1}^R {M_S}_{2,2}^R \nonumber \\
   &&+{M_{\rm mid}}_{(1,2)}^R ({M_S}_{1,2}^I)^2 -{M_{\rm mid}}_{(1,2)}^R ({M_S}_{1,2}^R)^2+{M_{\rm mid}}_{(2,2)}^I {M_S}_{1,1}^I {M_S}_{1,2}^R+{M_{\rm mid}}_{(2,2)}^I {M_S}_{1,1}^R {M_S}_{1,2}^I \nonumber \\
   &&+{M_{\rm mid}}_{(2,2)}^R {M_S}_{1,1}^I
   {M_S}_{1,2}^I -{M_{\rm mid}}_{(2,2)}^R {M_S}_{1,1}^R {M_S}_{1,2}^R)+{M_{\rm mid}}_{(1,2)}^I (2 {M_{\rm mid}}_{(1,2)}^R (-{M_S}_{2,2}^R ({M_{\rm mid}}_{(1,1)}^I {M_S}_{1,2}^R \nonumber \\
   &&+{M_{\rm mid}}_{(1,1)}^R {M_S}_{1,2}^I)+{M_{\rm mid}}_{(1,1)}^I {M_S}_{1,2}^I
   {M_S}_{2,2}^I-{M_{\rm mid}}_{(1,1)}^R {M_S}_{1,2}^R {M_S}_{2,2}^I+{M_{\rm mid}}_{(2,2)}^I {M_S}_{1,1}^I {M_S}_{1,2}^I \nonumber \\
   &&-{M_{\rm mid}}_{(2,2)}^I {M_S}_{1,1}^R {M_S}_{1,2}^R -{M_{\rm mid}}_{(2,2)}^R {M_S}_{1,1}^I {M_S}_{1,2}^R -{M_{\rm mid}}_{(2,2)}^R {M_S}_{1,1}^R
   {M_S}_{1,2}^I) \nonumber \\
   &&+{M_{\rm mid}}_{(1,1)}^I ({M_{\rm mid}}_{(2,2)}^R ({M_S}_{1,1}^I {M_S}_{2,2}^I -{M_S}_{1,1}^R {M_S}_{2,2}^R+({M_S}_{1,2}^I)^2-({M_S}_{1,2}^R)^2) \nonumber \\
  && -{M_{\rm mid}}_{(2,2)}^I ({M_S}_{1,1}^I {M_S}_{2,2}^R +{M_S}_{1,1}^R
   {M_S}_{2,2}^I+2 {M_S}_{1,2}^I {M_S}_{1,2}^R))+{M_{\rm mid}}_{(1,1)}^R ({M_{\rm mid}}_{(2,2)}^I ({M_S}_{1,1}^I {M_S}_{2,2}^I \nonumber \\
   && -{M_S}_{1,1}^R {M_S}_{2,2}^R+({M_S}_{1,2}^I)^2-({M_S}_{1,2}^R)^2)+{M_{\rm mid}}_{(2,2)}^R
   ({M_S}_{1,1}^I {M_S}_{2,2}^R+{M_S}_{1,1}^R {M_S}_{2,2}^I+2 {M_S}_{1,2}^I {M_S}_{1,2}^R)) \nonumber \\
   &&+({M_{\rm mid}}_{(1,2)}^R)^2 ({M_S}_{1,1}^I {M_S}_{2,2}^R +{M_S}_{1,1}^R {M_S}_{2,2}^I+2 {M_S}_{1,2}^I
   {M_S}_{1,2}^R))+{M_{\rm mid}}_{(1,1)}^I ({M_{\rm mid}}_{(1,2)}^R)^2 {M_S}_{1,2}^R {M_S}_{2,2}^I \nonumber \\
   &&-{M_{\rm mid}}_{(1,1)}^I {M_{\rm mid}}_{(1,2)}^R {M_{\rm mid}}_{(2,2)}^I {M_S}_{1,1}^I {M_S}_{2,2}^I-{M_{\rm mid}}_{(1,1)}^I {M_{\rm mid}}_{(1,2)}^R {M_{\rm mid}}_{(2,2)}^I
   ({M_S}_{1,2}^I)^2 \nonumber \\
   && +{M_{\rm mid}}_{(1,1)}^I {M_{\rm mid}}_{(1,2)}^R {M_{\rm mid}}_{(2,2)}^I ({M_S}_{1,2}^R)^2-{M_{\rm mid}}_{(1,1)}^I {M_{\rm mid}}_{(1,2)}^R {M_{\rm mid}}_{(2,2)}^R {M_S}_{1,1}^R {M_S}_{2,2}^I \nonumber \\
   && -2 {M_{\rm mid}}_{(1,1)}^I {M_{\rm mid}}_{(1,2)}^R {M_{\rm mid}}_{(2,2)}^R {M_S}_{1,2}^I
   {M_S}_{1,2}^R+{M_{\rm mid}}_{(1,1)}^I ({M_{\rm mid}}_{(2,2)}^I)^2 {M_S}_{1,1}^I {M_S}_{1,2}^R \nonumber \\
   &&+{M_{\rm mid}}_{(1,1)}^I ({M_{\rm mid}}_{(2,2)}^I)^2 {M_S}_{1,1}^R {M_S}_{1,2}^I+{M_{\rm mid}}_{(1,1)}^I ({M_{\rm mid}}_{(2,2)}^R)^2 {M_S}_{1,1}^I {M_S}_{1,2}^R \nonumber \\
   &&+{M_{\rm mid}}_{(1,1)}^I
   ({M_{\rm mid}}_{(2,2)}^R)^2 {M_S}_{1,1}^R {M_S}_{1,2}^I +({M_{\rm mid}}_{(1,1)}^R)^2 ({M_{\rm mid}}_{(2,2)}^I {M_S}_{1,2}^I {M_S}_{2,2}^R \nonumber \\
   &&+{M_{\rm mid}}_{(2,2)}^I {M_S}_{1,2}^R {M_S}_{2,2}^I -{M_{\rm mid}}_{(2,2)}^R {M_S}_{1,2}^I {M_S}_{2,2}^I+{M_{\rm mid}}_{(2,2)}^R
   {M_S}_{1,2}^R {M_S}_{2,2}^R) \nonumber \\
   &&-{M_{\rm mid}}_{(1,1)}^R (({M_{\rm mid}}_{(1,2)}^R)^2 ({M_S}_{1,2}^R {M_S}_{2,2}^R -{M_S}_{1,2}^I {M_S}_{2,2}^I) +{M_{\rm mid}}_{(1,2)}^R {M_{\rm mid}}_{(2,2)}^I ({M_S}_{1,1}^I {M_S}_{2,2}^R \nonumber \\
   &&+{M_S}_{1,1}^R {M_S}_{2,2}^I +2
   {M_S}_{1,2}^I {M_S}_{1,2}^R)+{M_{\rm mid}}_{(1,2)}^R {M_{\rm mid}}_{(2,2)}^R (-{M_S}_{1,1}^I {M_S}_{2,2}^I \nonumber \\
   &&+{M_S}_{1,1}^R {M_S}_{2,2}^R-({M_S}_{1,2}^I)^2+({M_S}_{1,2}^R)^2)+({M_{\rm mid}}_{(2,2)}^I)^2 ({M_S}_{1,1}^I
   {M_S}_{1,2}^I-{M_S}_{1,1}^R {M_S}_{1,2}^R) \nonumber \\
   &&+({M_{\rm mid}}_{(2,2)}^R)^2 ({M_S}_{1,1}^I {M_S}_{1,2}^I-{M_S}_{1,1}^R {M_S}_{1,2}^R))+({M_{\rm mid}}_{(1,2)}^I)^3 ({M_S}_{1,1}^I {M_S}_{2,2}^R+{M_S}_{1,1}^R {M_S}_{2,2}^I \nonumber \\
   &&+2 {M_S}_{1,2}^I
   {M_S}_{1,2}^R) -({M_{\rm mid}}_{(1,2)}^R)^3 {M_S}_{1,1}^I {M_S}_{2,2}^I-({M_{\rm mid}}_{(1,2)}^R)^3 ({M_S}_{1,2}^I)^2+({M_{\rm mid}}_{(1,2)}^R)^3 ({M_S}_{1,2}^R)^2 \nonumber \\
   &&+({M_{\rm mid}}_{(1,2)}^R)^2 {M_{\rm mid}}_{(2,2)}^I {M_S}_{1,1}^I {M_S}_{1,2}^R +({M_{\rm mid}}_{(1,2)}^R)^2
   {M_{\rm mid}}_{(2,2)}^I {M_S}_{1,1}^R {M_S}_{1,2}^I \nonumber \\
   &&+({M_{\rm mid}}_{(1,2)}^R)^2 {M_{\rm mid}}_{(2,2)}^R {M_S}_{1,1}^I {M_S}_{1,2}^I-({M_{\rm mid}}_{(1,2)}^R)^2 {M_{\rm mid}}_{(2,2)}^R {M_S}_{1,1}^R {M_S}_{1,2}^R \,  
\eea
\bea
   B_{1,2}^R &=& (({M_{\rm mid}}_{(1,1)}^I)^2+({M_{\rm mid}}_{(1,1)}^R)^2)
   (({M_{\rm mid}}_{(2,2)}^I)^2+({M_{\rm mid}}_{(2,2)}^R)^2)+2 ({M_{\rm mid}}_{(1,2)}^I)^2 (-{M_{\rm mid}}_{(1,1)}^I {M_{\rm mid}}_{(2,2)}^I \nonumber \\
   && +{M_{\rm mid}}_{(1,1)}^R {M_{\rm mid}}_{(2,2)}^R+({M_{\rm mid}}_{(1,2)}^R)^2)-4 {M_{\rm mid}}_{(1,2)}^I {M_{\rm mid}}_{(1,2)}^R ({M_{\rm mid}}_{(1,1)}^I
   {M_{\rm mid}}_{(2,2)}^R \nonumber \\
   &&+{M_{\rm mid}}_{(1,1)}^R {M_{\rm mid}}_{(2,2)}^I) +2 ({M_{\rm mid}}_{(1,2)}^R)^2 ({M_{\rm mid}}_{(1,1)}^I {M_{\rm mid}}_{(2,2)}^I-{M_{\rm mid}}_{(1,1)}^R {M_{\rm mid}}_{(2,2)}^R)+({M_{\rm mid}}_{(1,2)}^I)^4 \nonumber \\
   &&+({M_{\rm mid}}_{(1,2)}^R)^4 \, 
   \eea
\bea
A_{1,2}^I &=&  {M_S}_{2,2}^R (({M_{\rm mid}}_{(1,1)}^I)^2+({M_{\rm mid}}_{(1,1)}^R)^2) ({M_{\rm mid}}_{(2,2)}^R {M_S}_{1,2}^I-{M_{\rm mid}}_{(2,2)}^I {M_S}_{1,2}^R) \nonumber \\
&& +({M_{\rm mid}}_{(1,1)}^I)^2 {M_{\rm mid}}_{(2,2)}^I {M_S}_{1,2}^I {M_S}_{2,2}^I+({M_{\rm mid}}_{(1,1)}^I)^2 {M_{\rm mid}}_{(2,2)}^R
   {M_S}_{1,2}^R {M_S}_{2,2}^I \nonumber \\
   &&+({M_{\rm mid}}_{(1,2)}^I)^2 (-{M_{\rm mid}}_{(1,1)}^I {M_S}_{1,2}^I {M_S}_{2,2}^I+{M_{\rm mid}}_{(1,1)}^I {M_S}_{1,2}^R {M_S}_{2,2}^R+{M_{\rm mid}}_{(1,1)}^R {M_S}_{1,2}^I {M_S}_{2,2}^R \nonumber \\
   &&+{M_{\rm mid}}_{(1,1)}^R {M_S}_{1,2}^R
   {M_S}_{2,2}^I+{M_{\rm mid}}_{(1,2)}^R {M_S}_{1,1}^I {M_S}_{2,2}^R+{M_{\rm mid}}_{(1,2)}^R {M_S}_{1,1}^R {M_S}_{2,2}^I \nonumber \\
   &&+2 {M_{\rm mid}}_{(1,2)}^R {M_S}_{1,2}^I {M_S}_{1,2}^R -{M_{\rm mid}}_{(2,2)}^I {M_S}_{1,1}^I {M_S}_{1,2}^I+{M_{\rm mid}}_{(2,2)}^I {M_S}_{1,1}^R
   {M_S}_{1,2}^R \nonumber \\
   &&+{M_{\rm mid}}_{(2,2)}^R {M_S}_{1,1}^I {M_S}_{1,2}^R +{M_{\rm mid}}_{(2,2)}^R {M_S}_{1,1}^R {M_S}_{1,2}^I) \nonumber \\
   &&+{M_{\rm mid}}_{(1,2)}^I (-2 {M_{\rm mid}}_{(1,2)}^R ({M_{\rm mid}}_{(1,1)}^I {M_S}_{1,2}^I {M_S}_{2,2}^R +{M_{\rm mid}}_{(1,1)}^I {M_S}_{1,2}^R
   {M_S}_{2,2}^I \nonumber \\
   &&+{M_{\rm mid}}_{(1,1)}^R {M_S}_{1,2}^I {M_S}_{2,2}^I-{M_{\rm mid}}_{(1,1)}^R {M_S}_{1,2}^R {M_S}_{2,2}^R +{M_{\rm mid}}_{(2,2)}^I {M_S}_{1,1}^I {M_S}_{1,2}^R \nonumber \\
   &&+{M_{\rm mid}}_{(2,2)}^I {M_S}_{1,1}^R {M_S}_{1,2}^I+{M_{\rm mid}}_{(2,2)}^R {M_S}_{1,1}^I
   {M_S}_{1,2}^I-{M_{\rm mid}}_{(2,2)}^R {M_S}_{1,1}^R {M_S}_{1,2}^R)\nonumber \\
   &&-{M_{\rm mid}}_{(1,1)}^I ({M_{\rm mid}}_{(2,2)}^I ({M_S}_{1,1}^I {M_S}_{2,2}^I-{M_S}_{1,1}^R {M_S}_{2,2}^R+({M_S}_{1,2}^I)^2-({M_S}_{1,2}^R)^2) \nonumber \\
   &&+{M_{\rm mid}}_{(2,2)}^R
   ({M_S}_{1,1}^I {M_S}_{2,2}^R +{M_S}_{1,1}^R {M_S}_{2,2}^I+2 {M_S}_{1,2}^I {M_S}_{1,2}^R)) \nonumber \\
   &&+{M_{\rm mid}}_{(1,1)}^R ({M_{\rm mid}}_{(2,2)}^R ({M_S}_{1,1}^I {M_S}_{2,2}^I-{M_S}_{1,1}^R
   {M_S}_{2,2}^R+({M_S}_{1,2}^I)^2 \nonumber \\
   && -({M_S}_{1,2}^R)^2)-{M_{\rm mid}}_{(2,2)}^I ({M_S}_{1,1}^I {M_S}_{2,2}^R+{M_S}_{1,1}^R {M_S}_{2,2}^I+2 {M_S}_{1,2}^I {M_S}_{1,2}^R)) \nonumber \\
   &&+({M_{\rm mid}}_{(1,2)}^R)^2 ({M_S}_{1,1}^I
   {M_S}_{2,2}^I -{M_S}_{1,1}^R {M_S}_{2,2}^R+({M_S}_{1,2}^I)^2-({M_S}_{1,2}^R)^2)) \nonumber \\
   &&+({M_{\rm mid}}_{(1,2)}^R)^2 (-{M_S}_{2,2}^R ({M_{\rm mid}}_{(1,1)}^I {M_S}_{1,2}^R +{M_{\rm mid}}_{(1,1)}^R {M_S}_{1,2}^I) +{M_{\rm mid}}_{(1,1)}^I {M_S}_{1,2}^I
   {M_S}_{2,2}^I \nonumber \\
   &&-{M_{\rm mid}}_{(1,1)}^R {M_S}_{1,2}^R {M_S}_{2,2}^I +{M_{\rm mid}}_{(2,2)}^I {M_S}_{1,1}^I {M_S}_{1,2}^I -{M_{\rm mid}}_{(2,2)}^I {M_S}_{1,1}^R {M_S}_{1,2}^R \nonumber \\
   &&-{M_{\rm mid}}_{(2,2)}^R {M_S}_{1,1}^I {M_S}_{1,2}^R-{M_{\rm mid}}_{(2,2)}^R {M_S}_{1,1}^R
   {M_S}_{1,2}^I) \nonumber \\
   &&+{M_{\rm mid}}_{(1,2)}^R ({M_{\rm mid}}_{(1,1)}^I ({M_{\rm mid}}_{(2,2)}^I ({M_S}_{1,1}^I {M_S}_{2,2}^R+{M_S}_{1,1}^R {M_S}_{2,2}^I \nonumber \\
   &&+2 {M_S}_{1,2}^I {M_S}_{1,2}^R)+{M_{\rm mid}}_{(2,2)}^R (-{M_S}_{1,1}^I
   {M_S}_{2,2}^I+{M_S}_{1,1}^R {M_S}_{2,2}^R \nonumber \\
   &&-({M_S}_{1,2}^I)^2+({M_S}_{1,2}^R)^2))-{M_{\rm mid}}_{(1,1)}^R ({M_{\rm mid}}_{(2,2)}^I ({M_S}_{1,1}^I {M_S}_{2,2}^I-{M_S}_{1,1}^R
   {M_S}_{2,2}^R \nonumber \\
   &&+({M_S}_{1,2}^I)^2-({M_S}_{1,2}^R)^2)+{M_{\rm mid}}_{(2,2)}^R ({M_S}_{1,1}^I {M_S}_{2,2}^R+{M_S}_{1,1}^R {M_S}_{2,2}^I+2 {M_S}_{1,2}^I {M_S}_{1,2}^R))) \nonumber \\
   &&+{M_{\rm mid}}_{(1,1)}^I ({M_{\rm mid}}_{(2,2)}^I)^2 {M_S}_{1,1}^I
   {M_S}_{1,2}^I-{M_{\rm mid}}_{(1,1)}^I ({M_{\rm mid}}_{(2,2)}^I)^2 {M_S}_{1,1}^R {M_S}_{1,2}^R \nonumber \\
   &&+{M_{\rm mid}}_{(1,1)}^I ({M_{\rm mid}}_{(2,2)}^R)^2 {M_S}_{1,1}^I {M_S}_{1,2}^I-{M_{\rm mid}}_{(1,1)}^I ({M_{\rm mid}}_{(2,2)}^R)^2 {M_S}_{1,1}^R {M_S}_{1,2}^R \nonumber \\
   &&+({M_{\rm mid}}_{(1,1)}^R)^2
   {M_{\rm mid}}_{(2,2)}^I {M_S}_{1,2}^I {M_S}_{2,2}^I+({M_{\rm mid}}_{(1,1)}^R)^2 {M_{\rm mid}}_{(2,2)}^R {M_S}_{1,2}^R {M_S}_{2,2}^I \nonumber \\
   &&+{M_{\rm mid}}_{(1,1)}^R ({M_{\rm mid}}_{(2,2)}^I)^2 {M_S}_{1,1}^I {M_S}_{1,2}^R+{M_{\rm mid}}_{(1,1)}^R ({M_{\rm mid}}_{(2,2)}^I)^2 {M_S}_{1,1}^R
   {M_S}_{1,2}^I \nonumber \\
   &&+{M_{\rm mid}}_{(1,1)}^R ({M_{\rm mid}}_{(2,2)}^R)^2 {M_S}_{1,1}^I {M_S}_{1,2}^R+{M_{\rm mid}}_{(1,1)}^R ({M_{\rm mid}}_{(2,2)}^R)^2 {M_S}_{1,1}^R {M_S}_{1,2}^I \nonumber \\
   &&+({M_{\rm mid}}_{(1,2)}^I)^3 ({M_S}_{1,1}^I {M_S}_{2,2}^I-{M_S}_{1,1}^R
   {M_S}_{2,2}^R+({M_S}_{1,2}^I)^2 \nonumber \\
   &&-({M_S}_{1,2}^R)^2)+({M_{\rm mid}}_{(1,2)}^R)^3 ({M_S}_{1,1}^I {M_S}_{2,2}^R+{M_S}_{1,1}^R {M_S}_{2,2}^I+2 {M_S}_{1,2}^I {M_S}_{1,2}^R) 
\eea
\bea
B_{1,2}^I &=& (({M_{\rm mid}}_{(1,1)}^I)^2+({M_{\rm mid}}_{(1,1)}^R)^2)
   (({M_{\rm mid}}_{(2,2)}^I)^2+({M_{\rm mid}}_{(2,2)}^R)^2)+2 ({M_{\rm mid}}_{(1,2)}^I)^2 (-{M_{\rm mid}}_{(1,1)}^I {M_{\rm mid}}_{(2,2)}^I \nonumber \\
   && +{M_{\rm mid}}_{(1,1)}^R {M_{\rm mid}}_{(2,2)}^R+({M_{\rm mid}}_{(1,2)}^R)^2)-4 {M_{\rm mid}}_{(1,2)}^I {M_{\rm mid}}_{(1,2)}^R ({M_{\rm mid}}_{(1,1)}^I
   {M_{\rm mid}}_{(2,2)}^R \nonumber \\
   && +{M_{\rm mid}}_{(1,1)}^R {M_{\rm mid}}_{(2,2)}^I) +2 ({M_{\rm mid}}_{(1,2)}^R)^2 ({M_{\rm mid}}_{(1,1)}^I {M_{\rm mid}}_{(2,2)}^I-{M_{\rm mid}}_{(1,1)}^R {M_{\rm mid}}_{(2,2)}^R)+({M_{\rm mid}}_{(1,2)}^I)^4 \nonumber \\
   &&+({M_{\rm mid}}_{(1,2)}^R)^4
\eea
\bea
A_{2,2}^R &=& 2 {M_S}_{2,2}^R ({M_S}_{2,2}^I ({M_{\rm mid}}_{(2,2)}^I (({M_{\rm mid}}_{(1,1)}^I)^2+({M_{\rm mid}}_{(1,1)}^R)^2)+{M_{\rm mid}}_{(1,1)}^I ({M_{\rm mid}}_{(1,2)}^R)^2) \nonumber \\
&& +{M_{\rm mid}}_{(1,2)}^R {M_S}_{1,2}^R ({M_{\rm mid}}_{(1,1)}^I
   {M_{\rm mid}}_{(2,2)}^I-{M_{\rm mid}}_{(1,1)}^R {M_{\rm mid}}_{(2,2)}^R+({M_{\rm mid}}_{(1,2)}^R)^2) \nonumber \\ 
   &&-{M_{\rm mid}}_{(1,2)}^R {M_S}_{1,2}^I ({M_{\rm mid}}_{(1,1)}^I {M_{\rm mid}}_{(2,2)}^R+{M_{\rm mid}}_{(1,1)}^R {M_{\rm mid}}_{(2,2)}^I)) \nonumber \\
   &&+({M_S}_{2,2}^R)^2 ({M_{\rm mid}}_{(2,2)}^R
   (({M_{\rm mid}}_{(1,1)}^I)^2+({M_{\rm mid}}_{(1,1)}^R)^2)-{M_{\rm mid}}_{(1,1)}^R ({M_{\rm mid}}_{(1,2)}^R)^2) \nonumber \\
   &&-({M_{\rm mid}}_{(1,1)}^I)^2 {M_{\rm mid}}_{(2,2)}^R ({M_S}_{2,2}^I)^2-({M_{\rm mid}}_{(1,2)}^I)^2 (2 {M_{\rm mid}}_{(1,1)}^I {M_S}_{2,2}^I
   {M_S}_{2,2}^R+{M_{\rm mid}}_{(1,1)}^R ({M_S}_{2,2}^I)^2 \nonumber \\
   &&-{M_{\rm mid}}_{(1,1)}^R ({M_S}_{2,2}^R)^2+2 {M_{\rm mid}}_{(1,2)}^R {M_S}_{1,2}^I {M_S}_{2,2}^I-2 {M_{\rm mid}}_{(1,2)}^R {M_S}_{1,2}^R {M_S}_{2,2}^R+2 {M_{\rm mid}}_{(2,2)}^I {M_S}_{1,2}^I
   {M_S}_{1,2}^R \nonumber \\
   &&+{M_{\rm mid}}_{(2,2)}^R ({M_S}_{1,2}^I-{M_S}_{1,2}^R) ({M_S}_{1,2}^I+{M_S}_{1,2}^R))+2 {M_{\rm mid}}_{(1,2)}^I ({M_{\rm mid}}_{(1,2)}^R ({M_{\rm mid}}_{(1,1)}^I ({M_S}_{2,2}^I \nonumber \\
   &&-{M_S}_{2,2}^R) ({M_S}_{2,2}^I+{M_S}_{2,2}^R)-2
   {M_{\rm mid}}_{(1,1)}^R {M_S}_{2,2}^I {M_S}_{2,2}^R+{M_{\rm mid}}_{(2,2)}^I ({M_S}_{1,2}^I-{M_S}_{1,2}^R) ({M_S}_{1,2}^I \nonumber \\
   &&+{M_S}_{1,2}^R)-2 {M_{\rm mid}}_{(2,2)}^R {M_S}_{1,2}^I {M_S}_{1,2}^R)+{M_S}_{2,2}^I (-{M_{\rm mid}}_{(1,1)}^I {M_{\rm mid}}_{(2,2)}^I
   {M_S}_{1,2}^R \nonumber \\
   &&+{M_{\rm mid}}_{(1,1)}^I {M_{\rm mid}}_{(2,2)}^R {M_S}_{1,2}^I+{M_{\rm mid}}_{(1,1)}^R {M_{\rm mid}}_{(2,2)}^I {M_S}_{1,2}^I+{M_{\rm mid}}_{(1,1)}^R {M_{\rm mid}}_{(2,2)}^R {M_S}_{1,2}^R) \nonumber \\
   &&-{M_S}_{2,2}^R ({M_{\rm mid}}_{(1,1)}^I {M_{\rm mid}}_{(2,2)}^I
   {M_S}_{1,2}^I+{M_{\rm mid}}_{(1,1)}^I {M_{\rm mid}}_{(2,2)}^R {M_S}_{1,2}^R+{M_{\rm mid}}_{(1,1)}^R {M_{\rm mid}}_{(2,2)}^I {M_S}_{1,2}^R \nonumber \\
   &&-{M_{\rm mid}}_{(1,1)}^R {M_{\rm mid}}_{(2,2)}^R {M_S}_{1,2}^I)+({M_{\rm mid}}_{(1,2)}^R)^2 ({M_S}_{1,2}^I {M_S}_{2,2}^R+{M_S}_{1,2}^R
   {M_S}_{2,2}^I)) \nonumber \\
   &&-2 {M_{\rm mid}}_{(1,1)}^I {M_{\rm mid}}_{(1,2)}^R {M_{\rm mid}}_{(2,2)}^I {M_S}_{1,2}^I {M_S}_{2,2}^I-2 {M_{\rm mid}}_{(1,1)}^I {M_{\rm mid}}_{(1,2)}^R {M_{\rm mid}}_{(2,2)}^R {M_S}_{1,2}^R {M_S}_{2,2}^I \nonumber \\
   &&+2 {M_{\rm mid}}_{(1,1)}^I ({M_{\rm mid}}_{(2,2)}^I)^2
   {M_S}_{1,2}^I {M_S}_{1,2}^R+2 {M_{\rm mid}}_{(1,1)}^I ({M_{\rm mid}}_{(2,2)}^R)^2 {M_S}_{1,2}^I {M_S}_{1,2}^R \nonumber \\
   &&-({M_{\rm mid}}_{(1,1)}^R)^2 {M_{\rm mid}}_{(2,2)}^R ({M_S}_{2,2}^I)^2+{M_{\rm mid}}_{(1,1)}^R ({M_{\rm mid}}_{(1,2)}^R)^2 ({M_S}_{2,2}^I)^2 \nonumber \\
   &&-2 {M_{\rm mid}}_{(1,1)}^R
   {M_{\rm mid}}_{(1,2)}^R {M_{\rm mid}}_{(2,2)}^I {M_S}_{1,2}^R {M_S}_{2,2}^I +2 {M_{\rm mid}}_{(1,1)}^R {M_{\rm mid}}_{(1,2)}^R {M_{\rm mid}}_{(2,2)}^R {M_S}_{1,2}^I {M_S}_{2,2}^I \nonumber \\
   &&-{M_{\rm mid}}_{(1,1)}^R ({M_{\rm mid}}_{(2,2)}^I)^2 ({M_S}_{1,2}^I)^2+{M_{\rm mid}}_{(1,1)}^R ({M_{\rm mid}}_{(2,2)}^I)^2
   ({M_S}_{1,2}^R)^2 \nonumber \\
   &&-{M_{\rm mid}}_{(1,1)}^R ({M_{\rm mid}}_{(2,2)}^R)^2 ({M_S}_{1,2}^I)^2+{M_{\rm mid}}_{(1,1)}^R ({M_{\rm mid}}_{(2,2)}^R)^2 ({M_S}_{1,2}^R)^2+2 ({M_{\rm mid}}_{(1,2)}^I)^3 ({M_S}_{1,2}^I {M_S}_{2,2}^R \nonumber \\
   &&+{M_S}_{1,2}^R {M_S}_{2,2}^I)-2 ({M_{\rm mid}}_{(1,2)}^R)^3
   {M_S}_{1,2}^I {M_S}_{2,2}^I+2 ({M_{\rm mid}}_{(1,2)}^R)^2 {M_{\rm mid}}_{(2,2)}^I {M_S}_{1,2}^I {M_S}_{1,2}^R \nonumber \\
   &&+({M_{\rm mid}}_{(1,2)}^R)^2 {M_{\rm mid}}_{(2,2)}^R ({M_S}_{1,2}^I)^2-({M_{\rm mid}}_{(1,2)}^R)^2 {M_{\rm mid}}_{(2,2)}^R
   ({M_S}_{1,2}^R)^2
\eea
\bea
B_{2,2}^R &=& (({M_{\rm mid}}_{(1,1)}^I)^2+({M_{\rm mid}}_{(1,1)}^R)^2) (({M_{\rm mid}}_{(2,2)}^I)^2+({M_{\rm mid}}_{(2,2)}^R)^2)+2 ({M_{\rm mid}}_{(1,2)}^I)^2 (-{M_{\rm mid}}_{(1,1)}^I {M_{\rm mid}}_{(2,2)}^I \nonumber \\
&& +{M_{\rm mid}}_{(1,1)}^R
   {M_{\rm mid}}_{(2,2)}^R+({M_{\rm mid}}_{(1,2)}^R)^2)-4 {M_{\rm mid}}_{(1,2)}^I {M_{\rm mid}}_{(1,2)}^R ({M_{\rm mid}}_{(1,1)}^I {M_{\rm mid}}_{(2,2)}^R \nonumber \\
   && +{M_{\rm mid}}_{(1,1)}^R {M_{\rm mid}}_{(2,2)}^I)+2 ({M_{\rm mid}}_{(1,2)}^R)^2 ({M_{\rm mid}}_{(1,1)}^I {M_{\rm mid}}_{(2,2)}^I-{M_{\rm mid}}_{(1,1)}^R
   {M_{\rm mid}}_{(2,2)}^R)+({M_{\rm mid}}_{(1,2)}^I)^4 \nonumber \\
   &&+({M_{\rm mid}}_{(1,2)}^R)^4
\eea
\bea
A_{2,2}^I &=& -{M_{\rm mid}}_{(2,2)}^I ({M_S}_{2,2}^R)^2 (({M_{\rm mid}}_{(1,1)}^I)^2+({M_{\rm mid}}_{(1,1)}^R)^2)+2 {M_{\rm mid}}_{(2,2)}^R {M_S}_{2,2}^I {M_S}_{2,2}^R (({M_{\rm mid}}_{(1,1)}^I)^2 \nonumber \\
&&+({M_{\rm mid}}_{(1,1)}^R)^2)+({M_{\rm mid}}_{(1,1)}^I)^2 {M_{\rm mid}}_{(2,2)}^I
   ({M_S}_{2,2}^I)^2+({M_{\rm mid}}_{(1,2)}^I)^2 (-{M_{\rm mid}}_{(1,1)}^I ({M_S}_{2,2}^I)^2 \nonumber \\
   &&+{M_{\rm mid}}_{(1,1)}^I ({M_S}_{2,2}^R)^2+2 {M_{\rm mid}}_{(1,1)}^R {M_S}_{2,2}^I {M_S}_{2,2}^R+2 {M_{\rm mid}}_{(1,2)}^R {M_S}_{1,2}^I {M_S}_{2,2}^R+2 {M_{\rm mid}}_{(1,2)}^R
   {M_S}_{1,2}^R {M_S}_{2,2}^I \nonumber \\
   &&+{M_{\rm mid}}_{(2,2)}^I (({M_S}_{1,2}^R)^2-({M_S}_{1,2}^I)^2)+2 {M_{\rm mid}}_{(2,2)}^R {M_S}_{1,2}^I {M_S}_{1,2}^R) \nonumber \\
   &&-2 {M_{\rm mid}}_{(1,2)}^I ({M_{\rm mid}}_{(1,2)}^R (2 {M_{\rm mid}}_{(1,1)}^I {M_S}_{2,2}^I
   {M_S}_{2,2}^R +{M_{\rm mid}}_{(1,1)}^R ({M_S}_{2,2}^I-{M_S}_{2,2}^R) ({M_S}_{2,2}^I+{M_S}_{2,2}^R) \nonumber \\
   &&+2 {M_{\rm mid}}_{(2,2)}^I {M_S}_{1,2}^I {M_S}_{1,2}^R+{M_{\rm mid}}_{(2,2)}^R ({M_S}_{1,2}^I-{M_S}_{1,2}^R)
   ({M_S}_{1,2}^I+{M_S}_{1,2}^R))\nonumber \\
   &&+{M_S}_{2,2}^I ({M_{\rm mid}}_{(1,1)}^I {M_{\rm mid}}_{(2,2)}^I {M_S}_{1,2}^I+{M_{\rm mid}}_{(1,1)}^I {M_{\rm mid}}_{(2,2)}^R {M_S}_{1,2}^R+{M_{\rm mid}}_{(1,1)}^R {M_{\rm mid}}_{(2,2)}^I {M_S}_{1,2}^R \nonumber \\
   &&-{M_{\rm mid}}_{(1,1)}^R {M_{\rm mid}}_{(2,2)}^R
   {M_S}_{1,2}^I)+{M_S}_{2,2}^R (-{M_{\rm mid}}_{(1,1)}^I {M_{\rm mid}}_{(2,2)}^I {M_S}_{1,2}^R+{M_{\rm mid}}_{(1,1)}^I {M_{\rm mid}}_{(2,2)}^R {M_S}_{1,2}^I\nonumber \\
   &&+{M_{\rm mid}}_{(1,1)}^R {M_{\rm mid}}_{(2,2)}^I {M_S}_{1,2}^I+{M_{\rm mid}}_{(1,1)}^R {M_{\rm mid}}_{(2,2)}^R
   {M_S}_{1,2}^R) +({M_{\rm mid}}_{(1,2)}^R)^2 ({M_S}_{1,2}^R {M_S}_{2,2}^R -{M_S}_{1,2}^I {M_S}_{2,2}^I)) \nonumber \\
   &&+({M_{\rm mid}}_{(1,2)}^R)^2 ({M_{\rm mid}}_{(1,1)}^I ({M_S}_{2,2}^I-{M_S}_{2,2}^R) ({M_S}_{2,2}^I+{M_S}_{2,2}^R)-2 {M_{\rm mid}}_{(1,1)}^R
   {M_S}_{2,2}^I {M_S}_{2,2}^R \nonumber \\
   &&+{M_{\rm mid}}_{(2,2)}^I ({M_S}_{1,2}^I-{M_S}_{1,2}^R) ({M_S}_{1,2}^I+{M_S}_{1,2}^R)-2 {M_{\rm mid}}_{(2,2)}^R {M_S}_{1,2}^I {M_S}_{1,2}^R) \nonumber \\
   &&-2 {M_{\rm mid}}_{(1,2)}^R {M_S}_{2,2}^I (-{M_{\rm mid}}_{(1,1)}^I {M_{\rm mid}}_{(2,2)}^I
   {M_S}_{1,2}^R+{M_{\rm mid}}_{(1,1)}^I {M_{\rm mid}}_{(2,2)}^R {M_S}_{1,2}^I\nonumber \\
   &&+{M_{\rm mid}}_{(1,1)}^R {M_{\rm mid}}_{(2,2)}^I {M_S}_{1,2}^I+{M_{\rm mid}}_{(1,1)}^R {M_{\rm mid}}_{(2,2)}^R {M_S}_{1,2}^R)+2 {M_{\rm mid}}_{(1,2)}^R {M_S}_{2,2}^R ({M_{\rm mid}}_{(1,1)}^I {M_{\rm mid}}_{(2,2)}^I
   {M_S}_{1,2}^I \nonumber \\
   &&+{M_{\rm mid}}_{(1,1)}^I {M_{\rm mid}}_{(2,2)}^R {M_S}_{1,2}^R+{M_{\rm mid}}_{(1,1)}^R {M_{\rm mid}}_{(2,2)}^I {M_S}_{1,2}^R-{M_{\rm mid}}_{(1,1)}^R {M_{\rm mid}}_{(2,2)}^R {M_S}_{1,2}^I) \nonumber \\
   &&+{M_{\rm mid}}_{(1,1)}^I ({M_{\rm mid}}_{(2,2)}^I)^2 ({M_S}_{1,2}^I)^2-{M_{\rm mid}}_{(1,1)}^I
   ({M_{\rm mid}}_{(2,2)}^I)^2 ({M_S}_{1,2}^R)^2 \nonumber \\
   &&+{M_{\rm mid}}_{(1,1)}^I ({M_{\rm mid}}_{(2,2)}^R)^2 ({M_S}_{1,2}^I)^2-{M_{\rm mid}}_{(1,1)}^I ({M_{\rm mid}}_{(2,2)}^R)^2 ({M_S}_{1,2}^R)^2+({M_{\rm mid}}_{(1,1)}^R)^2 {M_{\rm mid}}_{(2,2)}^I ({M_S}_{2,2}^I)^2 \nonumber \\
   &&+2 {M_{\rm mid}}_{(1,1)}^R ({M_{\rm mid}}_{(2,2)}^I)^2
   {M_S}_{1,2}^I {M_S}_{1,2}^R+2 {M_{\rm mid}}_{(1,1)}^R ({M_{\rm mid}}_{(2,2)}^R)^2 {M_S}_{1,2}^I {M_S}_{1,2}^R \nonumber \\
   &&+2 ({M_{\rm mid}}_{(1,2)}^I)^3 ({M_S}_{1,2}^I {M_S}_{2,2}^I-{M_S}_{1,2}^R {M_S}_{2,2}^R)+2 ({M_{\rm mid}}_{(1,2)}^R)^3 ({M_S}_{1,2}^I
   {M_S}_{2,2}^R+{M_S}_{1,2}^R {M_S}_{2,2}^I)
\eea
\bea
B_{2,2}^I &=& (({M_{\rm mid}}_{(1,1)}^I)^2+({M_{\rm mid}}_{(1,1)}^R)^2) (({M_{\rm mid}}_{(2,2)}^I)^2+({M_{\rm mid}}_{(2,2)}^R)^2)+2 ({M_{\rm mid}}_{(1,2)}^I)^2 (-{M_{\rm mid}}_{(1,1)}^I {M_{\rm mid}}_{(2,2)}^I \nonumber \\
&& +{M_{\rm mid}}_{(1,1)}^R
   {M_{\rm mid}}_{(2,2)}^R+({M_{\rm mid}}_{(1,2)}^R)^2)-4 {M_{\rm mid}}_{(1,2)}^I {M_{\rm mid}}_{(1,2)}^R ({M_{\rm mid}}_{(1,1)}^I {M_{\rm mid}}_{(2,2)}^R \nonumber \\
   && +{M_{\rm mid}}_{(1,1)}^R {M_{\rm mid}}_{(2,2)}^I)+2 ({M_{\rm mid}}_{(1,2)}^R)^2 ({M_{\rm mid}}_{(1,1)}^I {M_{\rm mid}}_{(2,2)}^I-{M_{\rm mid}}_{(1,1)}^R
   {M_{\rm mid}}_{(2,2)}^R)+({M_{\rm mid}}_{(1,2)}^I)^4 \nonumber \\
   &&+({M_{\rm mid}}_{(1,2)}^R)^4
\eea
\section{Cross sections}
\label{app:B}
Expression for $\gamma_{D}^{(i)}$ can be written as \cite{Plumacher:1996kc} :
\bea
\gamma_{D}^{(i)} = N_{\tilde{\Psi}_i}^{eq}~ \frac{K_{1}(\frac{M_{\tilde{\Psi}_i}}{T})}{K_{2}(\frac{M_{\tilde{\Psi}_i}}{T})}~ \Gamma_{\tilde{\Psi_i}} \,, {\rm with}~ i=2 
\eea
$N_{\tilde{\Psi}_i}^{eq}$ being the equilibrium number density of mass eigenstate $\tilde{\Psi}_i$. Here $K_1$ and $K_2$ are the first and second modified Bessel functions of second kind respectively and $\Gamma_{\tilde{\Psi}_i}$ is the total decay width of $\tilde{\Psi}_i$.

Decay width of $\tilde{\Psi}_i$ at tree level, 
\bea
\Gamma_{\tilde{\Psi}_i} &:= & \Gamma (\tilde{\Psi}_i \rightarrow \phi^{\dagger} + l) + \Gamma (\tilde{\Psi}_i \rightarrow \phi + \bar{l}) \nonumber \\
&& = \frac{\alpha}{\text{sin}^{2}\theta_W} \frac{M_{\tilde{\Psi}_i}}{4} \frac{(M^{\dagger}_{D} M_{D})_{ii}}{M^{2}_{W}}
\label{decal-width}
\eea
with $\alpha, \theta_W$ being the Fine structure constant and the Weinberg angle.

For two body scattering $ a + b \rightarrow i + j + ...$, $\gamma_{eq}$ can be written as \cite{Plumacher:1996kc}, 
\bea
\gamma_{eq} = \frac{T}{64 \pi^{4}} \int_{(M_a+M_b)^2}^{\infty} \hspace{1mm} ds ~\hat{\sigma}(s) \hspace{1mm} \sqrt{s} \hspace{1mm}  K_{1}(\frac{\sqrt{s}}{T}) \,.
\eea
where $s$ \footnote{Not to be confused with "$s$-channel" mentioned earlier.} is the square of center of mass energy and $\hat{\sigma}(s)$ is reduced cross section, which can be expressed in terms of actual cross section as \cite{Plumacher:1996kc} :
\bea
\hat{\sigma}(s) = \frac{8}{s}\left[(p_a.p_b)^2 - M_a^2 M_b^2\right] \sigma(s) \,,
\eea 
with $p_k$ and $M_k$ being three momentum and mass of particle $k$.

The reduced cross-section for $L$-violating $s$-channel process is \cite{Plumacher:1996kc},
\bea
\hat{\sigma}_{N,s}(s)&=  &\frac{\alpha^{2}}{\text{sin}^{4}\theta_W} \frac{2 \pi}{M^{4}_{W}}\frac{1}{x}  \Biggl\{ \sum_{j=1}^2 a_{j}(M^{\dagger}_{D} M_{D})_{jj}^2 \Big[ \frac{x}{a_{j}} + \frac{2 x}{D_{j}(x)} + \frac{x^{2}}{2 D^{2}_{j}(x)} \nonumber \\ && - \Big(1 + 2\frac{x + a_{j}}{D_{j}(x)}\Big) ~\text{ln}\Big(\frac{x + a_{j}}{a_{j}}\Big) \Big] + 2 \sqrt{a_{1} a_{2}} ~\text{Re} \Big[ (M^{\dagger}_{D} M_{D})_{12}^2 \Big] \Bigg[ \frac{x}{D_{1}(x)} \nonumber \\ && + \frac{x}{D_{2}(x)} + \frac{x^{2}}{2 D_{1}(x) D_{2}(x)} - \frac{(x + a_{1})(x + a_{1} - 2 a_{2} )}{D_{2}(x) (a_{1} - a_{2})} ~\text{ln}\Big(\frac{x + a_{1}}{a_{1}}\Big) \nonumber \\ && - \frac{(x + a_{2})(x + a_{2} - 2 a_{1} )}{D_{1}(x) (a_{2} - a_{1})} ~\text{ln}\Big(\frac{x + a_{2}}{a_{2}}\Big) \Bigg] \Biggr\}.
\label{n1-schannel}
\eea
where, $x := \frac{s}{M_{\tilde{\Psi_1}}^2}, ~ a_i := \frac{M_{\tilde{\Psi_i}}^2}{M_{\tilde{\Psi_1}}^2}, ~
\frac{1}{D_{j}(x)} := \frac{x - a_{j}}{(x-a_{j})^{2} + a_{j} c_{j}}$, with $c_{j} := \Big(\frac{\Gamma_{\tilde{\Psi_j}}}{M_{\tilde{\Psi_1}}}\Big)^{2}$

The reduced cross-section for $L$-violating $t$-channel process is
\bea 
\hat{\sigma}_{N,t} (s) &=  & \frac{2 \pi \alpha^2 }{M^{4}_{W} \text{sin}^{4}\theta} \Biggl\{ \sum_{j=1}^2 a_{j}(M^{\dagger}_{D} M_{D})_{jj}^2 \hspace{2mm} \Big[\frac{1}{2 a_j}~\frac{x}{x+a_{j}} + \frac{1}{x + 2 a_{j}} \text{ln}\Big(\frac{x + a_{j}}{a_{j}}\Big)\Big] \nonumber \\ && + ~\text{Re} \Big[ (M^{\dagger}_{D} M_{D})_{12}^2 \Big] \frac{\sqrt{a_{1} a_{2}}}{(a_1 - a_2) (x + a_1 + a_2)} \Big[ (x + 2 a_1) ~\text{ln}\Big(\frac{x + a_{2}}{a_{2}} \Big) \nonumber \\ && - (x + 2 a_2) ~\text{ln}\Big(\frac{x + a_{1}}{a_{1}} \Big) \Big] \Biggr\}.
\label{n1-tchannel}
\eea
The reduced cross-section for s-channel process $\tilde{\Psi_j} + l \rightarrow \bar{t} + q$ (mediated by $\phi$) is 

\bea
\hat{\sigma}^{j}_{\phi,s}(s) = \frac{3 \pi \alpha^{2} M^{2}_{t}}{M^{4}_{W} \text{sin}^{4}\theta_W} (M^{\dagger}_{D} M_{D})_{jj} \times \Big(\frac{x - a_{j}}{x}\Big)^{2}.
\label{higgs-schannel}
\eea

The reduced cross-section for $t$-channel process $\tilde{\Psi_j} + t \rightarrow \bar{l} + q$ (mediated by $\phi$) is \cite{Plumacher:1996kc},
\bea
\hat{\sigma}^{j}_{\phi,t}(s) = \frac{3 \pi \alpha^{2} M^{2}_{t}}{M^{4}_{W} \text{sin}^{4}\theta_W} (M^{\dagger}_{D} M_{D})_{jj} \times \Big[\frac{x - a_{j}}{x} + \frac{a_{j}}{x}~ \text{ln} \Big(\frac{x - a_{j} + y~'}{y~'}\Big)\Big],
\label{higgs-tchannel}
\eea
where $y~' = \frac{M^{2}_h} {M_{\tilde{\Psi_1}}^{2}}$. 
\section{D.O.F}
\label{app:C}

Fermionic D.O.F can be calculated as :
\bea
g_{\rm fermion} &=& g_{\rm quark} + g_{\rm lepton} + g_{\rm neutrino} + g_{\rm RH-neutrino} \nonumber \\
&=& (6\times 3 \times 2 \times 2)+(3 \times 2 \times 2)+(3 \times 2)+( 2 \times 2) = 94
\eea
Whereas bosonic D.O.F. is :
\bea
g_{\rm boson} &=& g_{\rm gluon} + g_{\rm weak} + g_{\rm photon} + g_{\rm Higgs} \nonumber \\
&=& (8 \times 2)+(3 \times 3)+ 2 + 1 = 28
\eea
For our model the total D.O.F turns out to be :
\bea
g_{eff}~( T > 174 ~{\rm GeV}) = 28 + \frac{7}{8} \times 94 = 110.25
\eea
\bibliographystyle{JHEP}
\bibliography{refer1.bib}
\end{document}